\documentclass[twocolumn,aps,prl,superscriptaddress,longbibliography]{revtex4-2}
\usepackage[colorlinks=true, citecolor=blue, urlcolor=blue, linkcolor=red]{hyperref}
\renewcommand{\section}[1]{{\par\it #1.~~}\ignorespaces}
\usepackage{amsmath,amssymb,tikz,scalerel}
\usetikzlibrary{svg.path}
\definecolor{orcidlogocol}{HTML}{A6CE39}
\tikzset{orcidlogo/.pic={
		\fill[orcidlogocol] svg{M256,128c0,70.7-57.3,128-128,128C57.3,256,0,198.7,0,128C0,57.3,57.3,0,128,0C198.7,0,256,57.3,256,128z};
		\fill[white] svg{M86.3,186.2H70.9V79.1h15.4v48.4V186.2z}
		svg{M108.9,79.1h41.6c39.6,0,57,28.3,57,53.6c0,27.5-21.5,53.6-56.8,53.6h-41.8V79.1z M124.3,172.4h24.5c34.9,0,42.9-26.5,42.9-39.7c0-21.5-13.7-39.7-43.7-39.7h-23.7V172.4z}
		svg{M88.7,56.8c0,5.5-4.5,10.1-10.1,10.1c-5.6,0-10.1-4.6-10.1-10.1c0-5.6,4.5-10.1,10.1-10.1C84.2,46.7,88.7,51.3,88.7,56.8z};}}
\newcommand\orcid[1]{\href{https://orcid.org/#1}{\mbox{\scalerel*{\begin{tikzpicture}[yscale=-1,transform shape]\pic{orcidlogo};\end{tikzpicture}}{|}}}}

\begin{document}
\title{Breakdown of boundary criticality and exotic topological semimetals in $\mathcal{PT}$-invariant systems}
\author{Hong Wu\orcid{0000-0003-3276-7823}}
\affiliation{School of Electronic Science and Engineering, Chongqing University of Posts and Telecommunications, Chongqing 400065, China}
\author{Jun-Hong An\orcid{0000-0002-3475-0729}}\email{anjhong@lzu.edu.cn}
\affiliation{School of Physical Science and Technology and Lanzhou Center for Theoretical Physics, Lanzhou University, Lanzhou 730000, China}
\affiliation{Key Laboratory of Quantum Theory and Applications of MoE, Key Laboratory of Theoretical Physics of Gansu Province and Gansu Provincial Research Center for Basic Disciplines of Quantum Physics, Lanzhou University, Lanzhou 730000, China}

\begin{abstract}
It was recently found that, going beyond the tenfold Altland-Zirnbauer symmetry classes and violating the bulk-boundary correspondence of the usual topological phases, $\mathcal{PT}$-invariant systems support a real Chern insulator with the so-called boundary criticality, which forbids the transition between different orders of topological phases accompanied by the touching of bulk bands. Here, we find that periodic driving can break the boundary criticality of a $\mathcal{PT}$-invariant system. Upon setting the system free from the boundary criticality, diverse first- and second-order topological phases absent in the static case are found in both the zero and $\pi/T$ modes. In contrast to the usual Floquet ones defined on stroboscopic dynamics, our topological phases can only be characterized by the dynamical process of a whole driving period. The application of our result in a three-dimensional $\mathcal{PT}$-invariant system permits us to discover exotic second-order Dirac and nodal-line semimetals with coexisting surface and hinge Fermi arcs. Enriching the family of topological phases in $\mathcal{PT}$-invariant systems, our result provides a useful way to explore novel topological phases.
\end{abstract}

\maketitle
\section{Introduction}
Symmetry plays a central role in classifying quantum phases. Conventional quantum phases are governed by Landau's symmetry-breaking theory, which forms a basic paradigm of condensed matter physics to discover new quantum matters. Going beyond this paradigm, topological phases do not have the accompanied symmetry breaking. They are signified by the formation of symmetry-protected boundary states, which are characterized by the topology of the bulk energy bands \cite{RevModPhys.83.1057,RevModPhys.88.021004,RevModPhys.82.3045}. Being called a bulk-boundary correspondence, this is an essential principle of the topological phases. The topological phases are generically classified into the celebrated tenfold Altland-Zirnbauer symmetry classes according to whether time-reversal, particle-hole, and chiral symmetries are possessed by the systems \cite{RevModPhys.83.1057,RevModPhys.88.021004,RevModPhys.82.3045}. Under this classification rule, rich topological phases have been discovered \cite{PhysRevLett.111.086803,PhysRevB.78.195424,PhysRevLett.110.076401,RevModPhys.90.015001,PhysRevX.5.031013,PhysRevLett.125.090502,PhysRevLett.108.266802,PhysRevLett.118.106406,PhysRevLett.114.225301,PhysRevLett.98.106803,PhysRevLett.108.140405,PhysRevB.84.235126}.

\begin{figure*}[tbp]
\centering
\includegraphics[width=.8\textwidth]{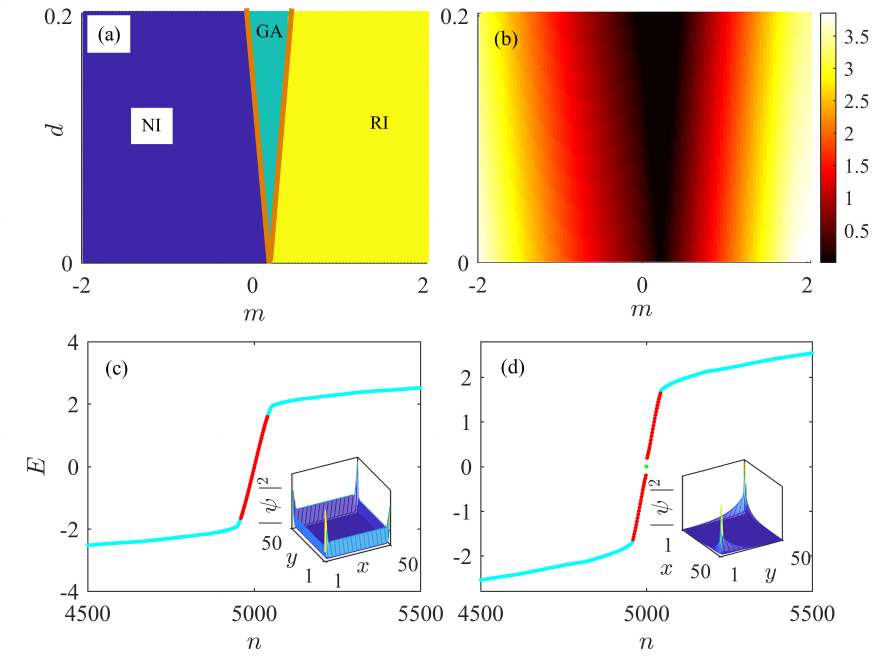}
\caption{(a) Phase diagram characterized by ${\mathcal V}_R$ and (b) bulk-energy gap in the $d$-$m$ plane. NI and RI denote a normal insulator with ${\mathcal V}_R=0$ and real Chern insulator with ${\mathcal V}_R=1$, respectively. GA denotes a gapless area with closed bulk-energy gap and thus ill-defined ${\mathcal V}_R$. Energy spectra under the open-boundary condition when (c) $d=0$ and (d) $0.2$. The blue and red lines are the bulk- and boundary-mode energies, respectively. The green point is the corner state. The insets show the probability distribution of the zero-mode edge state and corner states, respectively. We use $\lambda=1.3$ and $m=2$.} \label{linjie}
\end{figure*}

Recently, it was found that the above intrinsic symmetries do not exhaust the classes of topological phases. Various topological phases protected by spatial symmetries have been proposed \cite{PhysRevLett.129.046802,Benalcazar_2017,Schindler_2018,PhysRevX.9.011012,PhysRevLett.125.266804,PhysRevLett.125.146401,PhysRevB.105.L081102,Wei_2021,PhysRevLett.128.026405,PhysRevLett.127.177201,PhysRevLett.127.196801,PhysRevX.7.041069,PhysRevB.97.205135,shengjiehuang,PhysRevLett.126.156801,xiujuanzhang}. A distinguished example is that the systems with space-time inversion ($\mathcal{PT}$) symmetry exhibit diverse phases beyond the tenfold Altland-Zirnbauer symmetry classes \cite{PhysRevX.9.021013,PhysRevLett.123.256402,wang2024mirrorrealcherninsulator,RevModPhys.91.015006,PhysRevLett.114.114301,PhysRevA.109.053314,PhysRevB.107.085139,PhysRevX.8.031069,PhysRevB.106.L201406,PhysRevResearch.6.033192,PhysRevB.108.075129}. They can exhibit multigap topologies with band nodes that carry non-Abelian charges characterized by the braiding of such nodes \cite{Wu_2019,Guo_2021,PhysRevLett.132.216602,Jiang_2021,Bouhon_2020,pengbo}. Two-dimensional (2D) $\mathcal{PT}$-symmetric systems host a real Chern insulator \cite{jiang2022experimentalobservationmeronictopological,PhysRevLett.125.126403}. In contrast to the Chern insulator, which is defined in a single band with complex eigenfunctions of the Chern class, the real Chern insulator is defined in a set of bands with real eigenfunctions \cite{PhysRevLett.118.056401} and is characterized by the so-called real Chern number of the Stiefel-Whitney class \cite{PhysRevLett.121.106403,PhysRevB.95.235425,PhysRevB.109.195116,PhysRevX.8.031069}. Because the phase transition between the first- and second-order topological phases in a $\mathcal{P}\mathcal{T}$-invariant system originates from the closing and reopening of the boundary-mode energy bands rather than the bulk-mode ones, the first- and second-order topological phases in the real Chern insulator have the same real Chern number \cite{PhysRevLett.125.126403}. This is called boundary criticality and indicates that a real Chern insulator does not obey the usual bulk-boundary correspondence. Three-dimensional (3D) $\mathcal{PT}$-symmetric systems host either Dirac or nodal-line semimetals carrying a $\mathbb{Z}_2$ charge of the Stiefel-Whitney class, which is called a Stiefel-Whitney semimetal \cite{PhysRevLett.121.106403,PhysRevB.92.081201}. A Dirac-type Stiefel-Whitney semimetal is of first order and exhibits a surface Fermi arc, while a nodal-line-type one belongs to the second order and exhibits a hinge Fermi arc and drumhead surface states \cite{PhysRevLett.127.076401,PhysRevLett.126.196402}. The boundary criticality exerted by $\mathcal{PT}$ symmetry forbids the coexistence of surface and hinge Fermi arcs in a Stiefel-Whitney semimetal. From the perspective of application, one generally hopes for such a coexisting topological semimetal because it facilitates the utilization of the advantages of both the surface and hinge Fermi arcs. Thus, the boundary criticality, on the one hand, distinguishes the $\mathcal{PT}$-symmetric topological phases from the others, and on the other hand, it constrains the exploration of novel phases in such systems. 

Inspired by the progress that Floquet engineering has become a useful tool in creating exotic topological phases absent in static systems \cite{RevModPhys.89.011004,Bai_2021,PhysRevX.3.031005,floquetna,floquetshuyun,PhysRevLett.113.266801,PhysRevB.107.235132}, we propose a scheme to discover novel topological phases in a class of $\mathcal{P}\mathcal{T}$-invariant systems by applying a periodic driving. It is remarkable to find that the boundary criticality can be broken by periodic driving. Breaking through the constraint of the boundary criticality, a plethora of first- and second-order topological phases emerge, accompanied by the closing and reopening of the bulk-band gap, in both the zero and the $\pi/T$ modes. These phases are conspicuously absent in static $\mathcal{P}\mathcal{T}$-invariant systems. The further application of this result in a 3D $\mathcal{PT}$-symmetric system allows us to discover a variety of second-order topological semimetals characterized by the coexistence of surface and hinge Fermi arcs in both the Dirac and nodal-line types.

\section{Boundary criticality in a real Chern insulator}
We consider a spinless system in a 2D lattice whose Hamiltonian is $\mathcal{H}_0({\bf k})=\sum_i[(\cos k_i+\lambda)^2-\sin^2k_i-m]\Gamma_1+[2(\cos k_x+\lambda)\sin k_y-2(\cos k_y+\lambda)\sin k_x-m  ]\Gamma_2+[\sum_i 2(\cos k_i+\lambda)\sin k_i-m]\Gamma_3+id(\Gamma_2+\Gamma_3)\Gamma_4$, where $\Gamma_1=\sigma_0\tau_z$, $\Gamma_2=\sigma_y\tau_y$, $\Gamma_3=\sigma_0\tau_x$, $\Gamma_4=\sigma_x\tau_y$, and $\Gamma_5=\sigma_z\tau_y$. $\sigma_i$ and $\tau_i$ are two sets of Pauli matrices and $\sigma_0$ is the identity matrix.
It possesses $\mathcal{PT}$ symmetry, i.e., $\mathcal{PT}\mathcal{H}_0({\bf k})(\mathcal{PT})^{-1}=\mathcal{H}_0({\bf k})$, with $\mathcal{P}\mathcal{T}=\mathcal{K}$ denoting the complex conjugate, and the chiral symmetry $\Gamma_4\mathcal{H}_0({\bf k})\Gamma_4=-\mathcal{H}_0({\bf k})$. The topology of its bulk energy bands is characterized by the real Chern number \cite{PhysRevLett.125.126403}
\begin{equation}
\mathcal{V}_{R}=\int_{\rm{BZ}}\frac{d^2\mathbf{k}}{4\pi}\, \text{Tr} [I({\pmb\nabla}_{\bf{k}}\times {\mathcal{A}})_z]\mod 2,
\end{equation}
where ${\mathcal{A}}_{\beta\gamma}=\langle \beta,\bf{k} \lvert {\pmb\nabla}_{\bf{k}}  \lvert \gamma,\bf{k} \rangle$, $\lvert \beta/\gamma,\bf{k} \rangle$ are the real eigenstates of $\mathcal{H}_0({\bf k})$ under the reality requirement $\mathcal{P}\mathcal{T} \lvert  \beta/\gamma,\bf{k} \rangle=\lvert  \beta/\gamma,\bf{k} \rangle$, and $I=-i\sigma_y$ is the generator of the SO(2) group acting on the space spanned by the eigenstates of the two occupied bands. However, $\mathcal{V}_R=1$ cannot distinguish the first-order topological phase with gapless boundary states and the second-order one with gapped corner states. This is because the phase transition between different orders of topological phases in the $\mathcal{PT}$-symmetric system originates from the touching of the boundary-mode energy bands rather than the bulk-mode ones and thus cannot change $\mathcal{V}_R$. This is called boundary criticality \cite{PhysRevLett.125.126403}. Figure \ref{linjie}(a) shows a phase diagram characterized by $\mathcal{V}_R$. Except for the gapless regime with a closed bulk energy gap, where $\mathcal{V}_R$ is ill defined, the diagram splits into a normal insulator with $\mathcal{V}_R=0$ and a real Chern insulator with $\mathcal{V}_R=1$. The energy spectrum under the open-boundary condition in the $\mathcal{V}_R=1$ regime with $d=0$ confirms the first-order topological phases manifested by the gapless helical boundary modes [see Fig. \ref{linjie}(c)]. Without changing the bulk-energy topology and $\mathcal{V}_R$, the addition of a nonzero $d$ opens an energy gap of the boundary modes and leads to the formation of second-order topological phases manifested by the gapped corner states [see Fig. \ref{linjie}(d)]. Our system does not host a transition between different orders of topological phases induced by the closing and reopening of the bulk-energy gap in the static case. Therefore, the boundary criticality in a real Chern insulator exhibits a substantial difference from the bulk-boundary correspondence of the usual topological phases. 

\section{Boundary-criticality breakdown by periodic driving}
When the system is periodically driven, we have
\begin{equation}
\mathcal{H}({\bf k},t)=\mathcal{H}_1({\bf k})+\mathcal{H}_2({\bf k})\sum_{n\in\mathbb{Z}} \delta(t/T-n),\label{odr}
\end{equation}
where $\mathcal{H}_1({\bf k})$ is equal to $\mathcal{H}_0({\bf k})$ with $m=d=0$, $\mathcal{H}_2({\bf k})=(t_1\sin \alpha+t_2)\Gamma_1$, and $T$ is the driving period. Both $\mathcal{H}_1$ and $\mathcal{H}_2$ have $\mathcal{PT}$ symmetry. Because the energy of Eq. \eqref{odr} is not conserved, it does not have a well-defined energy spectrum. According to Floquet theorem, the one-period evolution operator ${U}(T)=\mathbb{T}e^{-i\int_{0}^{T}\mathcal{H}(t)dt}$ defines an effective Hamiltonian $\mathcal{H}_\text{eff}\equiv {i\over T}\ln {U}(T)$, whose eigenvalues are called quasienergies \cite{PhysRevA.91.052122,PhysRevLett.113.236803,PhysRevB.100.085138,PhysRevB.102.041119,PhysRevResearch.2.033494}. The topological phases of a time-periodic system are defined in the quasienergy spectrum \cite{PhysRevX.3.031005,PhysRevB.96.195303}. The original $\mathcal{P}\mathcal{T}$ symmetry is not inherited by $\mathcal{H}_\text{eff}({\bf k})$ due to $[\mathcal{H}_1,\mathcal{H}_2]\neq0$. After making a unitary transform $S=e^{-i\mathcal{H}_1({\bf k})T/2}$, we obtain $\mathcal{H}'_\text{eff}({\bf k})=\frac{i}{T}\ln[e^{-i\mathcal{H}_1({\bf k})T/2}e^{-i\mathcal{H}_2({\bf k})T}e^{-i\mathcal{H}_1({\bf k})T/2}]$, which shares the same quasienergy spectrum as $\mathcal{H}_\text{eff}({\bf k})$ but recovers the original $\mathcal{P}\mathcal{T}$ symmetry \cite{PhysRevB.104.205117}. $\mathcal{H}_\text{eff}'({\bf k})$ describes the stroboscopic dynamics of another periodic system,
\begin{equation}
\mathcal{H}'({\bf k},t) =\begin{cases} \mathcal{H}_1({\bf k}) ,& t\in\lbrack m, m+\frac{1}{2})T,\\ \mathcal{H}_2({\bf k}),& t\in\lbrack m+\frac{1}{2}, m+\frac{3}{2})T, \\
\mathcal{H}_1({\bf k}) ,& t\in\lbrack m+\frac{3}{2}, m+2)T,
\end{cases} \label{procotol}
\end{equation}
where $m$ is an even number. This lays the foundation for defining the topological invariants and revealing the bulk-boundary correspondence of our periodic system.

\begin{figure*}
\centering
\includegraphics[width=.8\textwidth]{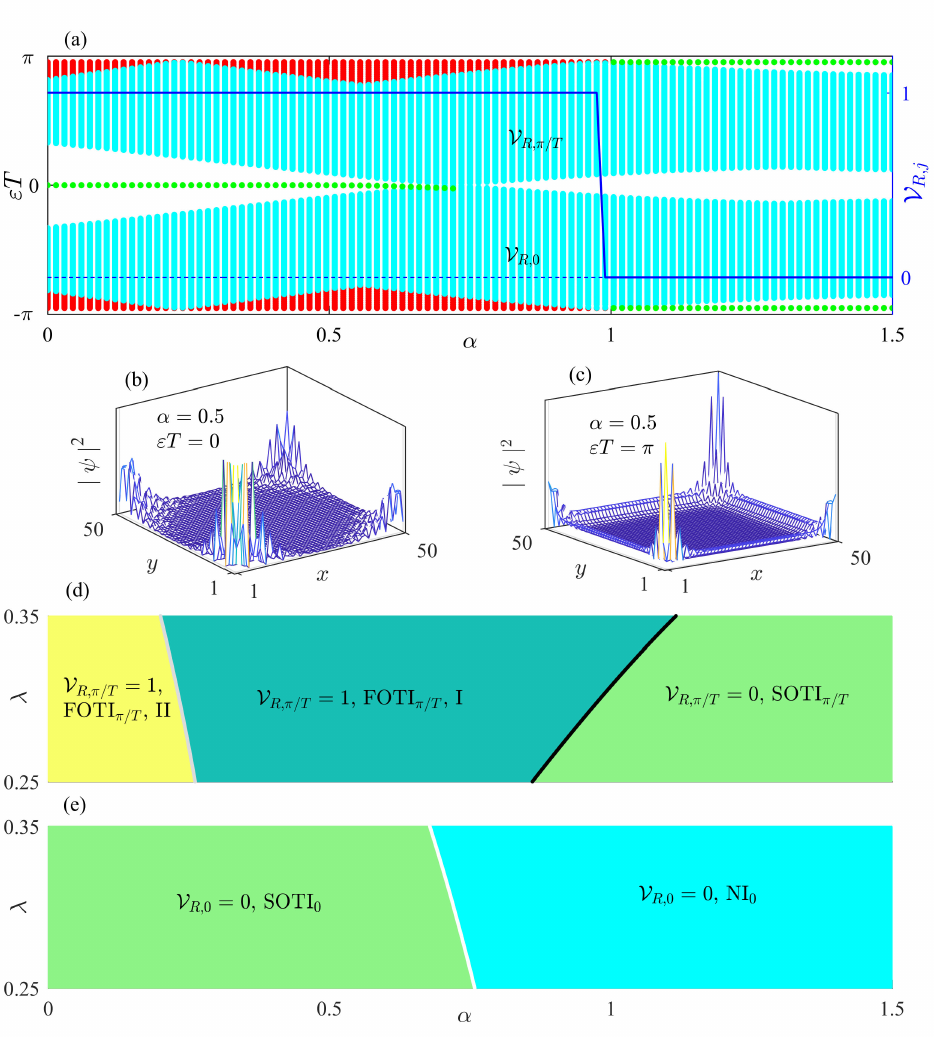}
\caption{ (a) Quasienergy spectra under open (red lines) and periodic (cyan lines) boundary conditions, $\mathcal{V}_{R,0}$, and $\mathcal{V}_{R,\pi/T}$ with the change of $\alpha$ when $\lambda=0.3$. The probability distribution of the topological (b) zero and (c) $\pi/T$ modes when $\alpha=0.5$. Phase diagram described by (d) $\mathcal{V}_{R,\pi/T}$ and (e) $\mathcal{V}_{R,0}$. SOTI is a second-order topological insulator, and two kinds of FOTI denote a first-order topological insulator whose gapless edge states cross at $k_y=\pi$ or $0$ in the cylinder geometry. We use $t_1=2$, $t_2=0.5$, $T=1$, and the size $40\times40$. } \label{energy}
\end{figure*}

A topological phase transition in a periodic system occurs not only at quasienergy zero but also at $\pi/T$. Thus, we need to define two topological invariants. This can be done in the following dynamical way. The diagonalization of the evolution operator
\begin{equation}
U'({\bf k},t) =\begin{cases} e^{-i\mathcal{H}_1t}, ~~~~~~~~~~~~~~~~~~~~~~t\in\lbrack 0, \frac{1}{2})T\\ e^{i\mathcal{H}_2(\frac{T}{2}-t)}e^{\frac{-i\mathcal{H}_1T}{2}},~~~~~~~~~t\in\lbrack \frac{1}{2},\frac{3}{2})T, \\
e^{i\mathcal{H}_1(\frac{3T}{2}-t)}e^{-i\mathcal{H}_2{T}}e^{\frac{-i\mathcal{H}_1T}{2}}, t\in\lbrack \frac{3}{2}, 2)T
\end{cases}
\end{equation}
where ${\bf k}$ in $\mathcal{H}_j({\bf k})$ has been omitted for brevity, yields $U'({\bf k},t)=\sum_s e^{-i\phi_s(\mathbf{k},t)}\lvert \psi_s(\mathbf{k},t) \rangle \langle \psi_s(\mathbf{k},t) \lvert$, where $\phi_s(\mathbf{k},t)=\varepsilon T$ forms a phase band \cite{zhang2020theoryanomalousfloquethigherorder}. During $t\in[0,2T]$, the phase bands may touch at zero or $\pi/T$ for some discrete points in the $\mathbf{k}$-$t$ space. This causes the difference between the topological invariants defined in $\phi_s(\mathbf{k},0)$ and $\phi_s(\mathbf{k},2T)$. Thus, the topological invariant at the quasienergy $\gamma/T$, with $\gamma=0$ or $\pi$, is
\begin{eqnarray}
\mathcal{V}_{R,\gamma/T}=\mathcal{V}_{R,\gamma/T}^{(0)}+\sum_{j}N_{j,\gamma}(\mathbf{k}_{j,\gamma},t_{j,\gamma}),
\end{eqnarray}
where $\mathcal{V}_{R,\gamma/T}^{(0)}$ is the real Chern number of $\mathcal{H}'({\bf k},0)$ and $N_{j,\gamma}$ is the topological charge of the $j$th band touching point $({\bf k}_{j,\gamma},t_{j,\gamma})$ making $\phi_s({\bf k}_{j,\gamma},t_{j,\gamma})=\gamma$. The topological charge is defined in the two lower phase bands as $N_{j,\gamma}=\oint_{{\mathcal S}_j}\frac{d\mathcal{S}_j}{4\pi}\, \text{Tr} [I({\pmb\nabla}_{\bf{k}}\times {\mathcal{A}})_z]\mod 2$, where ${\mathcal{A}}_{\beta\gamma}=\langle \psi_{\beta}({\bf k},t) \lvert {\pmb\nabla}_{\bf{k}}  \lvert \psi_{\gamma}({\bf k},t) \rangle$ and ${\mathcal S}_j$ is a small surface in ${\bf k}_{j,\gamma}$ enclosing $({\bf k}_{j,\gamma},t_{j,\gamma})$. Here, $\lvert \psi_{\beta/\gamma}({\bf k},t) \rangle$ are the eigenstates of the phase bands $\phi_{\beta/\gamma}({\bf k},t)\in(-\pi,0)$. This method gives a complete topological description of $\mathcal{PT}$-symmetric Floquet systems. Our topological invariants are defined in the phase bands within the dynamical process of a whole driving period, which endows our Floquet topological phases with a substantial difference from the usual ones defined only in the stroboscopic dynamics. 

It is expected that our periodic system, as a $\mathcal{PT}$-symmetric system, also holds the real Chern insulator with the boundary criticality. However, this is not true. We plot in Fig. \ref{energy}(a) the quasienergy spectrum in different $\alpha$. With the touching of the bulk bands at $\alpha=0.98$, a $\pi/T$-mode phase transition from the first- to the second-order topological phase occurs. Accompanied by an abrupt change of $\mathcal{V}_{R,\pi/T}$, this phase transition is witnessed by $\mathcal{V}_{R,\pi/T}$. This is in contrast to the boundary criticality in the static $\mathcal{PT}$-symmetric system in Fig. \ref{linjie}, where the transition between different orders of topological phases cannot change $\mathcal{V}_R$. It implies the breakdown of the boundary criticality induced by periodic driving. This behavior also occurs in the zero mode, where the appearance of a second-order topological phase is caused by the closing and reopening of the bulk-band gap instead of the boundary-band gap in the static case. The probability distributions in Figs. \ref{energy}(b) and \ref{energy}(c) when $\alpha=0.5$ confirm the second-order feature of the zero mode and the first-order feature of the $\pi/T$ mode. To gain a global picture of the topology of our periodic system, we plot in Figs. \ref{energy}(d) and \ref{energy}(e) the phase diagrams described by $\mathcal{V}_{R,\pi/T}$ and $\mathcal{V}_{R,0}$ in the $\lambda$-$\alpha$ plane. It shows that all the phase transitions originate from the bulk-band topology and thus the boundary criticality is completely absent in our periodically driven $\mathcal{PT}$-symmetric system.

Not existing in a static system, the second-order topological phase with a zero real Chern number in Figs. \ref{energy}(d) and \ref{energy}(e) is a distinctive feature of our periodically driven model.
The coexistence of first-order gapless boundary states at quasienergy $\pi/T$ with $\mathcal{V}_{R,\pi/T}=1$ and second-order gapped corner states the quasienergy zero with $\mathcal{V}_{R,\pi/T}=0$ also enriches the family of the real Chern insulator. However, as a $\mathbb{Z}_2$ topological invariant, $\mathcal{V}_{R,\gamma/T}$ fails to describe the second-order topological phases of our periodic system. To understand these exotic phases, we resort to the edge Hamiltonians \cite{PhysRevB.102.121405}. We first construct the edge Hamiltonian along the $x$ direction. Opening the boundary and performing the inverse Fourier transformation in the $x$ direction, we have $\sum_{\bf k}\hat{\Psi}^{\dag}_{\bf k}\mathcal{H}'_\text{eff}({\bf k })\hat{\Psi}_{{\bf k}}=\sum_{k_y}\sum_{j_x,j'_x} \hat{\Psi}^{\dag}_{k_y,j_x}{\mathsf H}_{j_x,j'_x}(k_y) \hat{\Psi}_{k_y,j'_x}$. The chiral symmetry enables ${\mathsf H}(k_y)$ to be rewritten in block-diagonal form as diag[${\mathsf H}^+(k_y)$,${\mathsf H}^-(k_y)$]. By solving the eigenequation ${\mathsf H}^\pm(k_y)\lvert \psi^{\pm}\rangle=\varepsilon^{\pm} \lvert \psi^{\pm} \rangle$, we obtain two left-edge states at the energy valley with $k_y=0$ as $\lvert \psi^\pm_{1} \rangle$ and $\lvert \psi^\pm_{2} \rangle$. The solution of $k_y\neq 0$ can be found by perturbatively expanding ${\mathsf H}^\pm(k_y)$ in the basis of $\lvert \psi^\pm_{1,2} \rangle$ as ${\mathsf H}^\pm(k_y)=\sum_{i,j}{\mathsf H}_{i,j}^{\pm,\text{L}}(k_y)|\psi_i^\pm\rangle\langle\psi_j^\pm|$. The left-edge Hamiltonian reads as ${\mathsf H}_{i,j}^{\pm,\text{L}}(k_y)=\langle\psi^\pm_i|{\mathsf H}^\pm(k_y) |\psi^\pm_j\rangle$. Similarly, the down-edge Hamiltonian ${\mathsf H}^{\pm,\text{D}}(k_x)$ along the $y$ direction can also be calculated. When $\alpha=1.18$, the numerical calculation shows ${\mathsf H}^{\pm,\text{L}}(k_y)=-2.1k_y\sigma_x\pm0.66k_y\sigma_y-1.81\sigma_z$ and ${\mathsf H}^{\pm,\text{D}}(k_x)=2.1k_x\sigma_x\pm0.66k_x\sigma_y+1.81\sigma_z$, which are a one-dimensional phases described by a $\mathbb{Z}_2$ topological invariant of the sign of the mass. Since the mass terms of ${\mathsf H}^{\pm,\text{L}}(k_y)$ and ${\mathsf H}^{\pm,\text{D}}(k_x)$ have opposite signs, they belong to distinct $\mathbb{Z}_2$ phases. According to the Jackiw-Rebbi theory \cite{PhysRevD.13.3398}, a 2D system formed by topologically nonequivalent neighboring one-dimensional systems supports a corner mode at the intersection of the two edges. This explains the origin of second-order topological phases in our system. Being not described by the real Chern number, such a $\mathcal{PT}$-symmetric second-order topological phase goes beyond the Stiefel-Whitney class.

\section{Exotic topological semimetals} 
Generalizing our system to the 3D case, we may create exotic topological semimetals, which can be sliced into a family of 2D $k_z$-dependent topological and normal phases \cite{PhysRevLett.125.266804,PhysRevLett.132.066601}. After replacing $m$ in $\mathcal{H}_0(\mathbf{k})$ by $2\cos k_z$, we obtain a static semimetal. We see from Fig. \ref{linjie}(a) that, when $d=0$, some 2D sliced systems are normal insulators and the others are first-order topological insulators. Thus, this 3D system is a first-order Dirac semimetal manifested as a surface Fermi arc. When $d\neq 0$, each Dirac point spreads into a nodal loop and the topologically nontrivial 2D system becomes a second-order topological insulator. Thus, the 3D system becomes a second-order nodal-line semimetal manifested as a hinge Fermi arc \cite{PhysRevLett.125.126403}. It also has drumhead surface states bounded by the projections of the nodal loops in the surface Brillouin zone characterized by the topological charge
\begin{equation}
w_{C}=\frac{1}{\pi}\oint_{C}d\mathbf{k}\cdot\text{Tr}[{\bf B}(\mathbf{k})],
\end{equation}
where $ C$ is a small circle transversely surrounding the nodal loop, ${\bf B}_{\beta\gamma}({\bf k})=\langle \beta,\mathbf{k}  \lvert  i{\pmb\nabla}_{\mathbf{k}} \lvert  \gamma,\mathbf{k} \rangle$ is the Berry connection given by smooth complex states, and $\lvert  \beta/\gamma,\mathbf{k} \rangle$ are the eigenstates of $\mathcal{H}'_\text{eff}$ \cite{PhysRevLett.125.126403}. $w_{C}$ of our system is $1$. Possessing hinge Fermi arcs and drumhead surface states, this second-order nodal-line semimetal has been observed in experiment \cite{PhysRevLett.132.197202}. Therefore, our static system is either a first-order Dirac semimetal or a second-order nodal-line semimetal and does not support the coexistence of a first-order surface Fermi arc and a second-order hinge Fermi arc and drumhead surface states due to the constraint of the boundary criticality.

\begin{figure*}[tbp]
\centering
\includegraphics[width=.8\textwidth]{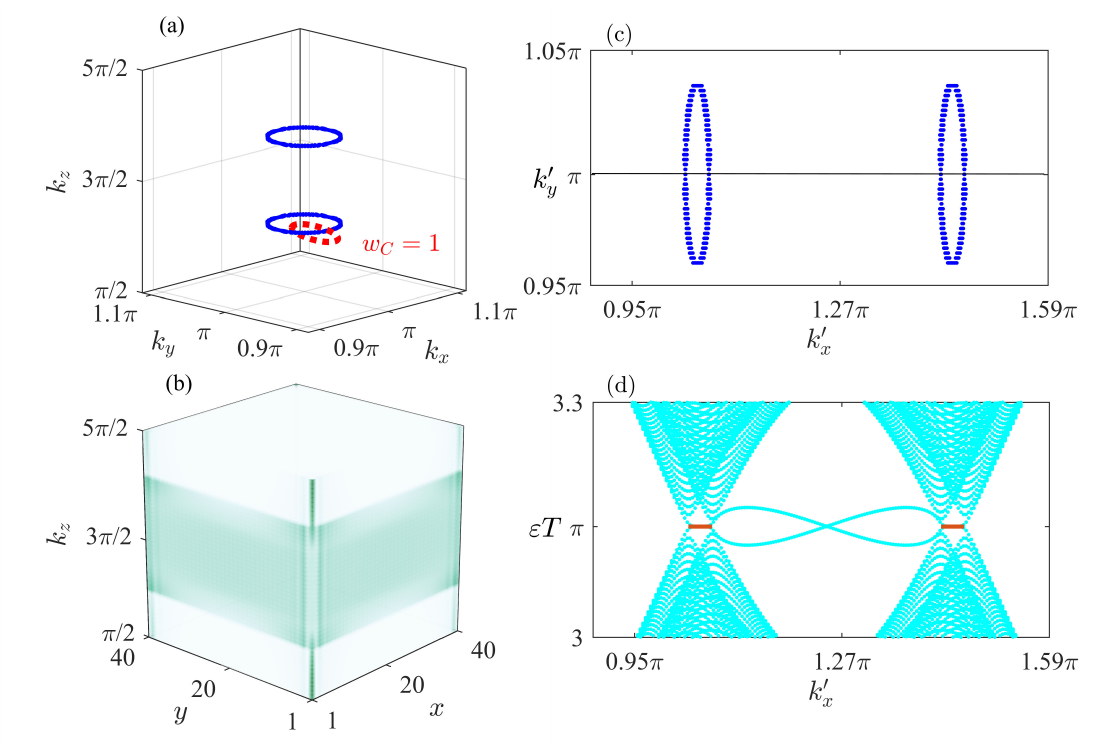}
\caption{(a) Nodal loops and (b) probability distributions of the $\pi/T$-mode states. (c) Nodal loops in momentum space with $k'_x=(k_z+k_x)/2$, $k'_y=k_y$, and $k'_z=(k_z-k_x)/2$. (d) Quasienergy spectrum under open boundary conditions along the $z'$ direction with $k'_y=\pi$. $z'$ represents the diagonal direction in the $x$-$z$ plane. The drumhead surface states are marked by red. We use $T=1$, $\lambda={0.3}$, $t_1=0.3$, $t_2=2.2$, and $m_{14}=0.5$.}
\label{sm}
\end{figure*}

Switching on the same periodic driving as Eq. \eqref{odr} but replacing $\alpha$ in $\mathcal{H}_2({\bf k})$ by $k_z$, we realize an exotic topological semimetal. First, it is exotic because it is a coexisting second-order topological semimetal at the quasienergies zero and $\pi/T$ [see Figs. \ref{energy}(b) and \ref{energy}(c)]. This cannot occur in the static case. Second, it is exotic because the $\pi/T$ mode supports the coexistence of a first-order surface Fermi arc and a second-order hinge Fermi arc. Although such a phase has been reported in Weyl phononic crystals \cite{Luo_2021}, it cannot occur in a $\mathcal{PT}$-symmetric system because the boundary criticality forbids a phase transition in its 2D sliced system between the first- and second-order phases caused by the touching of bulk energy bands. Our result shows that, by breaking the boundary criticality, Floquet engineering offers us a useful tool to create novel topological semimetals in $\mathcal{P}\mathcal{T}$-symmetric systems.

By further adding a perturbation $\Delta\mathcal{H}=im_{14}\Gamma_{1}\Gamma_4\sum_{n\in\mathbb{Z}} \delta(t/T-n)$ to our periodic system, each Dirac point spreads into a nodal loop [see Fig. \ref{sm}(a)]. The probability distributions of $\pi/T$-mode states in different $k_z$ reveals the coexisting surface and hinge Fermi arcs [see Fig. \ref{sm}(b)]. The drumhead surface states bounded by the projections of the nodal loop are captured by $w_{C}=1$. To confirm the drumhead surface states, we project the nodal loops in the plane spanned by $k'_x=\frac{k_z+k_x}{2}$ and $k'_y=k_y$ [see Fig. \ref{sm}(c)]. Opening the boundary along the direction $k'_z=\frac{k_z-k_x}{2}$, the quasienergy spectrum in Fig. \ref{sm}(d) clearly shows the appearance of two flat bands in the regimes enclosed by the nodal loops of Fig. \ref{sm}(c). They are just drumhead surface states. The result verifies that a $\mathcal{P}\mathcal{T}$-invariant nodal-line semimetal featuring coexisting surface, hinge Fermi arcs, and drumhead surface states is generated. This unique feature distinguishes the system from conventional higher-order nodal-line semimetals with only hinge Fermi arcs and drumhead surface states.  Although a similar coexistence of surface and hinge Fermi arcs has been found in Ref. \cite{PhysRevB.104.205117}, it requires a perturbation to break the time-reversal symmetry.  

\section{Discussion and conclusions}
It is noted that although only the delta-function driving protocol is considered, our results can be generalized to other forms. The 3D higher-order Stiefel-Whitney topological semimetals have been realized in $\mathcal{P}\mathcal{T}$-invariant sonic and photonic crystals \cite{Xue_2023,PhysRevLett.132.197202,Pan2023}. These systems have the same topology as our static system. On the other hand, periodic driving has exhibited its versatile power in engineering exotic phases in various platforms, such as ultracold atoms \cite{RevModPhys.89.011004,PhysRevLett.116.205301,PhysRevLett.130.043201}, superconductor qubits \cite{Roushan2017}, photonics \cite{Rechtsman2013,PhysRevLett.122.173901,Pan2023,PhysRevLett.133.073803}, and acoustic systems \cite{PhysRevLett.129.254301}. This progress gives strong support to the experimental realization of exotic $\mathcal{PT}$-symmetric topological phases. 

We have investigated the topological phases in periodically driven $\mathcal{P}\mathcal{T}$-invariant systems. It is found that exotic topological phase transitions between the first- and second-order topological insulators accompanied by the closing and reopening of the bulk-band gap, which are forbidden by the boundary criticality in static systems, are triggered by periodic driving. It reveals the breakdown of the boundary criticality by periodic driving. The generalization of this scheme to 3D $\mathcal{P}\mathcal{T}$-symmetric systems permits us to realize anomalous Dirac and nodal-line semimetals featuring coexisting surface Fermi arcs, hinge Fermi arcs, and drumhead surface states, which are forbidden by the boundary criticality in static systems. Our result reveals that, supplying an alternate dimension to manipulate different kinds of bulk-boundary correspondence, Floquet engineering opens an avenue for realizing exotic topological phases without static analogs.

\section{Acknowledgments}
The work is supported by the National Natural Science Foundation of China (Grants No. 12405007, No. 12275109, and No. 12247101), and the Innovation Program for Quantum Science and Technology of China (Grant No. 2023ZD0300904).

\bibliography{references}

\begin{thebibliography}{86}%
\makeatletter
\providecommand \@ifxundefined [1]{%
 \@ifx{#1\undefined}
}%
\providecommand \@ifnum [1]{%
 \ifnum #1\expandafter \@firstoftwo
 \else \expandafter \@secondoftwo
 \fi
}%
\providecommand \@ifx [1]{%
 \ifx #1\expandafter \@firstoftwo
 \else \expandafter \@secondoftwo
 \fi
}%
\providecommand \natexlab [1]{#1}%
\providecommand \enquote  [1]{``#1''}%
\providecommand \bibnamefont  [1]{#1}%
\providecommand \bibfnamefont [1]{#1}%
\providecommand \citenamefont [1]{#1}%
\providecommand \href@noop [0]{\@secondoftwo}%
\providecommand \href [0]{\begingroup \@sanitize@url \@href}%
\providecommand \@href[1]{\@@startlink{#1}\@@href}%
\providecommand \@@href[1]{\endgroup#1\@@endlink}%
\providecommand \@sanitize@url [0]{\catcode `\\12\catcode `\$12\catcode
  `\&12\catcode `\#12\catcode `\^12\catcode `\_12\catcode `\%12\relax}%
\providecommand \@@startlink[1]{}%
\providecommand \@@endlink[0]{}%
\providecommand \url  [0]{\begingroup\@sanitize@url \@url }%
\providecommand \@url [1]{\endgroup\@href {#1}{\urlprefix }}%
\providecommand \urlprefix  [0]{URL }%
\providecommand \Eprint [0]{\href }%
\providecommand \doibase [0]{https://doi.org/}%
\providecommand \selectlanguage [0]{\@gobble}%
\providecommand \bibinfo  [0]{\@secondoftwo}%
\providecommand \bibfield  [0]{\@secondoftwo}%
\providecommand \translation [1]{[#1]}%
\providecommand \BibitemOpen [0]{}%
\providecommand \bibitemStop [0]{}%
\providecommand \bibitemNoStop [0]{.\EOS\space}%
\providecommand \EOS [0]{\spacefactor3000\relax}%
\providecommand \BibitemShut  [1]{\csname bibitem#1\endcsname}%
\let\auto@bib@innerbib\@empty
\bibitem [{\citenamefont {Qi}\ and\ \citenamefont
  {Zhang}(2011)}]{RevModPhys.83.1057}%
  \BibitemOpen
  \bibfield  {author} {\bibinfo {author} {\bibfnamefont {X.-L.}\ \bibnamefont
  {Qi}}\ and\ \bibinfo {author} {\bibfnamefont {S.-C.}\ \bibnamefont {Zhang}},\
  }\bibfield  {title} {\bibinfo {title} {Topological insulators and
  superconductors},\ }\href {https://doi.org/10.1103/RevModPhys.83.1057}
  {\bibfield  {journal} {\bibinfo  {journal} {Rev. Mod. Phys.}\ }\textbf
  {\bibinfo {volume} {83}},\ \bibinfo {pages} {1057} (\bibinfo {year}
  {2011})}\BibitemShut {NoStop}%
\bibitem [{\citenamefont {Bansil}\ \emph {et~al.}(2016)\citenamefont {Bansil},
  \citenamefont {Lin},\ and\ \citenamefont {Das}}]{RevModPhys.88.021004}%
  \BibitemOpen
  \bibfield  {author} {\bibinfo {author} {\bibfnamefont {A.}~\bibnamefont
  {Bansil}}, \bibinfo {author} {\bibfnamefont {H.}~\bibnamefont {Lin}},\ and\
  \bibinfo {author} {\bibfnamefont {T.}~\bibnamefont {Das}},\ }\bibfield
  {title} {\bibinfo {title} {Colloquium: Topological band theory},\ }\href
  {https://doi.org/10.1103/RevModPhys.88.021004} {\bibfield  {journal}
  {\bibinfo  {journal} {Rev. Mod. Phys.}\ }\textbf {\bibinfo {volume} {88}},\
  \bibinfo {pages} {021004} (\bibinfo {year} {2016})}\BibitemShut {NoStop}%
\bibitem [{\citenamefont {Hasan}\ and\ \citenamefont
  {Kane}(2010)}]{RevModPhys.82.3045}%
  \BibitemOpen
  \bibfield  {author} {\bibinfo {author} {\bibfnamefont {M.~Z.}\ \bibnamefont
  {Hasan}}\ and\ \bibinfo {author} {\bibfnamefont {C.~L.}\ \bibnamefont
  {Kane}},\ }\bibfield  {title} {\bibinfo {title} {Colloquium: Topological
  insulators},\ }\href {https://doi.org/10.1103/RevModPhys.82.3045} {\bibfield
  {journal} {\bibinfo  {journal} {Rev. Mod. Phys.}\ }\textbf {\bibinfo {volume}
  {82}},\ \bibinfo {pages} {3045} (\bibinfo {year} {2010})}\BibitemShut
  {NoStop}%
\bibitem [{\citenamefont {Wang}\ \emph {et~al.}(2013)\citenamefont {Wang},
  \citenamefont {Lian}, \citenamefont {Zhang},\ and\ \citenamefont
  {Zhang}}]{PhysRevLett.111.086803}%
  \BibitemOpen
  \bibfield  {author} {\bibinfo {author} {\bibfnamefont {J.}~\bibnamefont
  {Wang}}, \bibinfo {author} {\bibfnamefont {B.}~\bibnamefont {Lian}}, \bibinfo
  {author} {\bibfnamefont {H.}~\bibnamefont {Zhang}},\ and\ \bibinfo {author}
  {\bibfnamefont {S.-C.}\ \bibnamefont {Zhang}},\ }\bibfield  {title} {\bibinfo
  {title} {Anomalous edge transport in the quantum anomalous {H}all state},\
  }\href {https://doi.org/10.1103/PhysRevLett.111.086803} {\bibfield  {journal}
  {\bibinfo  {journal} {Phys. Rev. Lett.}\ }\textbf {\bibinfo {volume} {111}},\
  \bibinfo {pages} {086803} (\bibinfo {year} {2013})}\BibitemShut {NoStop}%
\bibitem [{\citenamefont {Qi}\ \emph {et~al.}(2008)\citenamefont {Qi},
  \citenamefont {Hughes},\ and\ \citenamefont {Zhang}}]{PhysRevB.78.195424}%
  \BibitemOpen
  \bibfield  {author} {\bibinfo {author} {\bibfnamefont {X.-L.}\ \bibnamefont
  {Qi}}, \bibinfo {author} {\bibfnamefont {T.~L.}\ \bibnamefont {Hughes}},\
  and\ \bibinfo {author} {\bibfnamefont {S.-C.}\ \bibnamefont {Zhang}},\
  }\bibfield  {title} {\bibinfo {title} {Topological field theory of
  time-reversal invariant insulators},\ }\href
  {https://doi.org/10.1103/PhysRevB.78.195424} {\bibfield  {journal} {\bibinfo
  {journal} {Phys. Rev. B}\ }\textbf {\bibinfo {volume} {78}},\ \bibinfo
  {pages} {195424} (\bibinfo {year} {2008})}\BibitemShut {NoStop}%
\bibitem [{\citenamefont {Liu}\ \emph {et~al.}(2013)\citenamefont {Liu},
  \citenamefont {Liu},\ and\ \citenamefont {Cheng}}]{PhysRevLett.110.076401}%
  \BibitemOpen
  \bibfield  {author} {\bibinfo {author} {\bibfnamefont {X.-J.}\ \bibnamefont
  {Liu}}, \bibinfo {author} {\bibfnamefont {Z.-X.}\ \bibnamefont {Liu}},\ and\
  \bibinfo {author} {\bibfnamefont {M.}~\bibnamefont {Cheng}},\ }\bibfield
  {title} {\bibinfo {title} {Manipulating topological edge spins in a
  one-dimensional optical lattice},\ }\href
  {https://doi.org/10.1103/PhysRevLett.110.076401} {\bibfield  {journal}
  {\bibinfo  {journal} {Phys. Rev. Lett.}\ }\textbf {\bibinfo {volume} {110}},\
  \bibinfo {pages} {076401} (\bibinfo {year} {2013})}\BibitemShut {NoStop}%
\bibitem [{\citenamefont {Armitage}\ \emph {et~al.}(2018)\citenamefont
  {Armitage}, \citenamefont {Mele},\ and\ \citenamefont
  {Vishwanath}}]{RevModPhys.90.015001}%
  \BibitemOpen
  \bibfield  {author} {\bibinfo {author} {\bibfnamefont {N.~P.}\ \bibnamefont
  {Armitage}}, \bibinfo {author} {\bibfnamefont {E.~J.}\ \bibnamefont {Mele}},\
  and\ \bibinfo {author} {\bibfnamefont {A.}~\bibnamefont {Vishwanath}},\
  }\bibfield  {title} {\bibinfo {title} {Weyl and {D}irac semimetals in
  three-dimensional solids},\ }\href
  {https://doi.org/10.1103/RevModPhys.90.015001} {\bibfield  {journal}
  {\bibinfo  {journal} {Rev. Mod. Phys.}\ }\textbf {\bibinfo {volume} {90}},\
  \bibinfo {pages} {015001} (\bibinfo {year} {2018})}\BibitemShut {NoStop}%
\bibitem [{\citenamefont {Lv}\ \emph {et~al.}(2015)\citenamefont {Lv},
  \citenamefont {Weng}, \citenamefont {Fu}, \citenamefont {Wang}, \citenamefont
  {Miao}, \citenamefont {Ma}, \citenamefont {Richard}, \citenamefont {Huang},
  \citenamefont {Zhao}, \citenamefont {Chen}, \citenamefont {Fang},
  \citenamefont {Dai}, \citenamefont {Qian},\ and\ \citenamefont
  {Ding}}]{PhysRevX.5.031013}%
  \BibitemOpen
  \bibfield  {author} {\bibinfo {author} {\bibfnamefont {B.~Q.}\ \bibnamefont
  {Lv}}, \bibinfo {author} {\bibfnamefont {H.~M.}\ \bibnamefont {Weng}},
  \bibinfo {author} {\bibfnamefont {B.~B.}\ \bibnamefont {Fu}}, \bibinfo
  {author} {\bibfnamefont {X.~P.}\ \bibnamefont {Wang}}, \bibinfo {author}
  {\bibfnamefont {H.}~\bibnamefont {Miao}}, \bibinfo {author} {\bibfnamefont
  {J.}~\bibnamefont {Ma}}, \bibinfo {author} {\bibfnamefont {P.}~\bibnamefont
  {Richard}}, \bibinfo {author} {\bibfnamefont {X.~C.}\ \bibnamefont {Huang}},
  \bibinfo {author} {\bibfnamefont {L.~X.}\ \bibnamefont {Zhao}}, \bibinfo
  {author} {\bibfnamefont {G.~F.}\ \bibnamefont {Chen}}, \bibinfo {author}
  {\bibfnamefont {Z.}~\bibnamefont {Fang}}, \bibinfo {author} {\bibfnamefont
  {X.}~\bibnamefont {Dai}}, \bibinfo {author} {\bibfnamefont {T.}~\bibnamefont
  {Qian}},\ and\ \bibinfo {author} {\bibfnamefont {H.}~\bibnamefont {Ding}},\
  }\bibfield  {title} {\bibinfo {title} {Experimental discovery of {W}eyl
  semimetal {T}a{A}s},\ }\href {https://doi.org/10.1103/PhysRevX.5.031013}
  {\bibfield  {journal} {\bibinfo  {journal} {Phys. Rev. X}\ }\textbf {\bibinfo
  {volume} {5}},\ \bibinfo {pages} {031013} (\bibinfo {year}
  {2015})}\BibitemShut {NoStop}%
\bibitem [{\citenamefont {Xin}\ \emph {et~al.}(2020)\citenamefont {Xin},
  \citenamefont {Li}, \citenamefont {Fan}, \citenamefont {Zhu}, \citenamefont
  {Zhang}, \citenamefont {Nie}, \citenamefont {Li}, \citenamefont {Liu},\ and\
  \citenamefont {Lu}}]{PhysRevLett.125.090502}%
  \BibitemOpen
  \bibfield  {author} {\bibinfo {author} {\bibfnamefont {T.}~\bibnamefont
  {Xin}}, \bibinfo {author} {\bibfnamefont {Y.}~\bibnamefont {Li}}, \bibinfo
  {author} {\bibfnamefont {Y.-a.}\ \bibnamefont {Fan}}, \bibinfo {author}
  {\bibfnamefont {X.}~\bibnamefont {Zhu}}, \bibinfo {author} {\bibfnamefont
  {Y.}~\bibnamefont {Zhang}}, \bibinfo {author} {\bibfnamefont
  {X.}~\bibnamefont {Nie}}, \bibinfo {author} {\bibfnamefont {J.}~\bibnamefont
  {Li}}, \bibinfo {author} {\bibfnamefont {Q.}~\bibnamefont {Liu}},\ and\
  \bibinfo {author} {\bibfnamefont {D.}~\bibnamefont {Lu}},\ }\bibfield
  {title} {\bibinfo {title} {Quantum phases of three-dimensional chiral
  topological insulators on a spin quantum simulator},\ }\href
  {https://doi.org/10.1103/PhysRevLett.125.090502} {\bibfield  {journal}
  {\bibinfo  {journal} {Phys. Rev. Lett.}\ }\textbf {\bibinfo {volume} {125}},\
  \bibinfo {pages} {090502} (\bibinfo {year} {2020})}\BibitemShut {NoStop}%
\bibitem [{\citenamefont {Fang}\ \emph {et~al.}(2012)\citenamefont {Fang},
  \citenamefont {Gilbert}, \citenamefont {Dai},\ and\ \citenamefont
  {Bernevig}}]{PhysRevLett.108.266802}%
  \BibitemOpen
  \bibfield  {author} {\bibinfo {author} {\bibfnamefont {C.}~\bibnamefont
  {Fang}}, \bibinfo {author} {\bibfnamefont {M.~J.}\ \bibnamefont {Gilbert}},
  \bibinfo {author} {\bibfnamefont {X.}~\bibnamefont {Dai}},\ and\ \bibinfo
  {author} {\bibfnamefont {B.~A.}\ \bibnamefont {Bernevig}},\ }\bibfield
  {title} {\bibinfo {title} {Multi-{W}eyl topological semimetals stabilized by
  point group symmetry},\ }\href
  {https://doi.org/10.1103/PhysRevLett.108.266802} {\bibfield  {journal}
  {\bibinfo  {journal} {Phys. Rev. Lett.}\ }\textbf {\bibinfo {volume} {108}},\
  \bibinfo {pages} {266802} (\bibinfo {year} {2012})}\BibitemShut {NoStop}%
\bibitem [{\citenamefont {Xu}\ \emph {et~al.}(2017)\citenamefont {Xu},
  \citenamefont {Aut\`es}, \citenamefont {Matt}, \citenamefont {Lv},
  \citenamefont {Yao}, \citenamefont {Bisti}, \citenamefont {Strocov},
  \citenamefont {Gawryluk}, \citenamefont {Pomjakushina}, \citenamefont
  {Conder}, \citenamefont {Plumb}, \citenamefont {Radovic}, \citenamefont
  {Qian}, \citenamefont {Yazyev}, \citenamefont {Mesot}, \citenamefont {Ding},\
  and\ \citenamefont {Shi}}]{PhysRevLett.118.106406}%
  \BibitemOpen
  \bibfield  {author} {\bibinfo {author} {\bibfnamefont {N.}~\bibnamefont
  {Xu}}, \bibinfo {author} {\bibfnamefont {G.}~\bibnamefont {Aut\`es}},
  \bibinfo {author} {\bibfnamefont {C.~E.}\ \bibnamefont {Matt}}, \bibinfo
  {author} {\bibfnamefont {B.~Q.}\ \bibnamefont {Lv}}, \bibinfo {author}
  {\bibfnamefont {M.~Y.}\ \bibnamefont {Yao}}, \bibinfo {author} {\bibfnamefont
  {F.}~\bibnamefont {Bisti}}, \bibinfo {author} {\bibfnamefont {V.~N.}\
  \bibnamefont {Strocov}}, \bibinfo {author} {\bibfnamefont {D.}~\bibnamefont
  {Gawryluk}}, \bibinfo {author} {\bibfnamefont {E.}~\bibnamefont
  {Pomjakushina}}, \bibinfo {author} {\bibfnamefont {K.}~\bibnamefont
  {Conder}}, \bibinfo {author} {\bibfnamefont {N.~C.}\ \bibnamefont {Plumb}},
  \bibinfo {author} {\bibfnamefont {M.}~\bibnamefont {Radovic}}, \bibinfo
  {author} {\bibfnamefont {T.}~\bibnamefont {Qian}}, \bibinfo {author}
  {\bibfnamefont {O.~V.}\ \bibnamefont {Yazyev}}, \bibinfo {author}
  {\bibfnamefont {J.}~\bibnamefont {Mesot}}, \bibinfo {author} {\bibfnamefont
  {H.}~\bibnamefont {Ding}},\ and\ \bibinfo {author} {\bibfnamefont
  {M.}~\bibnamefont {Shi}},\ }\bibfield  {title} {\bibinfo {title} {Distinct
  evolutions of {W}eyl fermion quasiparticles and {F}ermi arcs with bulk band
  topology in {W}eyl semimetals},\ }\href
  {https://doi.org/10.1103/PhysRevLett.118.106406} {\bibfield  {journal}
  {\bibinfo  {journal} {Phys. Rev. Lett.}\ }\textbf {\bibinfo {volume} {118}},\
  \bibinfo {pages} {106406} (\bibinfo {year} {2017})}\BibitemShut {NoStop}%
\bibitem [{\citenamefont {Dub\ifmmode~\check{c}\else \v{c}\fi{}ek}\ \emph
  {et~al.}(2015)\citenamefont {Dub\ifmmode~\check{c}\else \v{c}\fi{}ek},
  \citenamefont {Kennedy}, \citenamefont {Lu}, \citenamefont {Ketterle},
  \citenamefont {Solja\ifmmode \check{c}\else
  \v{c}\fi{}i\ifmmode~\acute{c}\else \'{c}\fi{}},\ and\ \citenamefont
  {Buljan}}]{PhysRevLett.114.225301}%
  \BibitemOpen
  \bibfield  {author} {\bibinfo {author} {\bibfnamefont {T.}~\bibnamefont
  {Dub\ifmmode~\check{c}\else \v{c}\fi{}ek}}, \bibinfo {author} {\bibfnamefont
  {C.~J.}\ \bibnamefont {Kennedy}}, \bibinfo {author} {\bibfnamefont
  {L.}~\bibnamefont {Lu}}, \bibinfo {author} {\bibfnamefont {W.}~\bibnamefont
  {Ketterle}}, \bibinfo {author} {\bibfnamefont {M.}~\bibnamefont
  {Solja\ifmmode \check{c}\else \v{c}\fi{}i\ifmmode~\acute{c}\else
  \'{c}\fi{}}},\ and\ \bibinfo {author} {\bibfnamefont {H.}~\bibnamefont
  {Buljan}},\ }\bibfield  {title} {\bibinfo {title} {Weyl points in
  three-dimensional optical lattices: Synthetic magnetic monopoles in momentum
  space},\ }\href {https://doi.org/10.1103/PhysRevLett.114.225301} {\bibfield
  {journal} {\bibinfo  {journal} {Phys. Rev. Lett.}\ }\textbf {\bibinfo
  {volume} {114}},\ \bibinfo {pages} {225301} (\bibinfo {year}
  {2015})}\BibitemShut {NoStop}%
\bibitem [{\citenamefont {Fu}\ \emph {et~al.}(2007)\citenamefont {Fu},
  \citenamefont {Kane},\ and\ \citenamefont {Mele}}]{PhysRevLett.98.106803}%
  \BibitemOpen
  \bibfield  {author} {\bibinfo {author} {\bibfnamefont {L.}~\bibnamefont
  {Fu}}, \bibinfo {author} {\bibfnamefont {C.~L.}\ \bibnamefont {Kane}},\ and\
  \bibinfo {author} {\bibfnamefont {E.~J.}\ \bibnamefont {Mele}},\ }\bibfield
  {title} {\bibinfo {title} {Topological insulators in three dimensions},\
  }\href {https://doi.org/10.1103/PhysRevLett.98.106803} {\bibfield  {journal}
  {\bibinfo  {journal} {Phys. Rev. Lett.}\ }\textbf {\bibinfo {volume} {98}},\
  \bibinfo {pages} {106803} (\bibinfo {year} {2007})}\BibitemShut {NoStop}%
\bibitem [{\citenamefont {Young}\ \emph {et~al.}(2012)\citenamefont {Young},
  \citenamefont {Zaheer}, \citenamefont {Teo}, \citenamefont {Kane},
  \citenamefont {Mele},\ and\ \citenamefont {Rappe}}]{PhysRevLett.108.140405}%
  \BibitemOpen
  \bibfield  {author} {\bibinfo {author} {\bibfnamefont {S.~M.}\ \bibnamefont
  {Young}}, \bibinfo {author} {\bibfnamefont {S.}~\bibnamefont {Zaheer}},
  \bibinfo {author} {\bibfnamefont {J.~C.~Y.}\ \bibnamefont {Teo}}, \bibinfo
  {author} {\bibfnamefont {C.~L.}\ \bibnamefont {Kane}}, \bibinfo {author}
  {\bibfnamefont {E.~J.}\ \bibnamefont {Mele}},\ and\ \bibinfo {author}
  {\bibfnamefont {A.~M.}\ \bibnamefont {Rappe}},\ }\bibfield  {title} {\bibinfo
  {title} {Dirac semimetal in three dimensions},\ }\href
  {https://doi.org/10.1103/PhysRevLett.108.140405} {\bibfield  {journal}
  {\bibinfo  {journal} {Phys. Rev. Lett.}\ }\textbf {\bibinfo {volume} {108}},\
  \bibinfo {pages} {140405} (\bibinfo {year} {2012})}\BibitemShut {NoStop}%
\bibitem [{\citenamefont {Burkov}\ \emph {et~al.}(2011)\citenamefont {Burkov},
  \citenamefont {Hook},\ and\ \citenamefont {Balents}}]{PhysRevB.84.235126}%
  \BibitemOpen
  \bibfield  {author} {\bibinfo {author} {\bibfnamefont {A.~A.}\ \bibnamefont
  {Burkov}}, \bibinfo {author} {\bibfnamefont {M.~D.}\ \bibnamefont {Hook}},\
  and\ \bibinfo {author} {\bibfnamefont {L.}~\bibnamefont {Balents}},\
  }\bibfield  {title} {\bibinfo {title} {Topological nodal semimetals},\ }\href
  {https://doi.org/10.1103/PhysRevB.84.235126} {\bibfield  {journal} {\bibinfo
  {journal} {Phys. Rev. B}\ }\textbf {\bibinfo {volume} {84}},\ \bibinfo
  {pages} {235126} (\bibinfo {year} {2011})}\BibitemShut {NoStop}%
\bibitem [{\citenamefont {Tanaka}\ \emph {et~al.}(2022)\citenamefont {Tanaka},
  \citenamefont {Zhang}, \citenamefont {Uwaha},\ and\ \citenamefont
  {Murakami}}]{PhysRevLett.129.046802}%
  \BibitemOpen
  \bibfield  {author} {\bibinfo {author} {\bibfnamefont {Y.}~\bibnamefont
  {Tanaka}}, \bibinfo {author} {\bibfnamefont {T.}~\bibnamefont {Zhang}},
  \bibinfo {author} {\bibfnamefont {M.}~\bibnamefont {Uwaha}},\ and\ \bibinfo
  {author} {\bibfnamefont {S.}~\bibnamefont {Murakami}},\ }\bibfield  {title}
  {\bibinfo {title} {Anomalous crystal shapes of topological crystalline
  insulators},\ }\href {https://doi.org/10.1103/PhysRevLett.129.046802}
  {\bibfield  {journal} {\bibinfo  {journal} {Phys. Rev. Lett.}\ }\textbf
  {\bibinfo {volume} {129}},\ \bibinfo {pages} {046802} (\bibinfo {year}
  {2022})}\BibitemShut {NoStop}%
\bibitem [{\citenamefont {Benalcazar}\ \emph {et~al.}(2017)\citenamefont
  {Benalcazar}, \citenamefont {Bernevig},\ and\ \citenamefont
  {Hughes}}]{Benalcazar_2017}%
  \BibitemOpen
  \bibfield  {author} {\bibinfo {author} {\bibfnamefont {W.~A.}\ \bibnamefont
  {Benalcazar}}, \bibinfo {author} {\bibfnamefont {B.~A.}\ \bibnamefont
  {Bernevig}},\ and\ \bibinfo {author} {\bibfnamefont {T.~L.}\ \bibnamefont
  {Hughes}},\ }\bibfield  {title} {\bibinfo {title} {Quantized electric
  multipole insulators},\ }\href {https://doi.org/10.1126/science.aah6442}
  {\bibfield  {journal} {\bibinfo  {journal} {Science}\ }\textbf {\bibinfo
  {volume} {357}},\ \bibinfo {pages} {61} (\bibinfo {year} {2017})}\BibitemShut
  {NoStop}%
\bibitem [{\citenamefont {Schindler}\ \emph {et~al.}(2018)\citenamefont
  {Schindler}, \citenamefont {Cook}, \citenamefont {Vergniory}, \citenamefont
  {Wang}, \citenamefont {Parkin}, \citenamefont {Bernevig},\ and\ \citenamefont
  {Neupert}}]{Schindler_2018}%
  \BibitemOpen
  \bibfield  {author} {\bibinfo {author} {\bibfnamefont {F.}~\bibnamefont
  {Schindler}}, \bibinfo {author} {\bibfnamefont {A.~M.}\ \bibnamefont {Cook}},
  \bibinfo {author} {\bibfnamefont {M.~G.}\ \bibnamefont {Vergniory}}, \bibinfo
  {author} {\bibfnamefont {Z.}~\bibnamefont {Wang}}, \bibinfo {author}
  {\bibfnamefont {S.~S.~P.}\ \bibnamefont {Parkin}}, \bibinfo {author}
  {\bibfnamefont {B.~A.}\ \bibnamefont {Bernevig}},\ and\ \bibinfo {author}
  {\bibfnamefont {T.}~\bibnamefont {Neupert}},\ }\bibfield  {title} {\bibinfo
  {title} {Higher-order topological insulators},\ }\href
  {https://doi.org/10.1126/sciadv.aat0346} {\bibfield  {journal} {\bibinfo
  {journal} {Science Advances}\ }\textbf {\bibinfo {volume} {4}},\ \bibinfo
  {pages} {eaat0346} (\bibinfo {year} {2018})}\BibitemShut {NoStop}%
\bibitem [{\citenamefont {Trifunovic}\ and\ \citenamefont
  {Brouwer}(2019)}]{PhysRevX.9.011012}%
  \BibitemOpen
  \bibfield  {author} {\bibinfo {author} {\bibfnamefont {L.}~\bibnamefont
  {Trifunovic}}\ and\ \bibinfo {author} {\bibfnamefont {P.~W.}\ \bibnamefont
  {Brouwer}},\ }\bibfield  {title} {\bibinfo {title} {Higher-order
  bulk-boundary correspondence for topological crystalline phases},\ }\href
  {https://doi.org/10.1103/PhysRevX.9.011012} {\bibfield  {journal} {\bibinfo
  {journal} {Phys. Rev. X}\ }\textbf {\bibinfo {volume} {9}},\ \bibinfo {pages}
  {011012} (\bibinfo {year} {2019})}\BibitemShut {NoStop}%
\bibitem [{\citenamefont {Ghorashi}\ \emph {et~al.}(2020)\citenamefont
  {Ghorashi}, \citenamefont {Li},\ and\ \citenamefont
  {Hughes}}]{PhysRevLett.125.266804}%
  \BibitemOpen
  \bibfield  {author} {\bibinfo {author} {\bibfnamefont {S.~A.~A.}\
  \bibnamefont {Ghorashi}}, \bibinfo {author} {\bibfnamefont {T.}~\bibnamefont
  {Li}},\ and\ \bibinfo {author} {\bibfnamefont {T.~L.}\ \bibnamefont
  {Hughes}},\ }\bibfield  {title} {\bibinfo {title} {Higher-order {W}eyl
  semimetals},\ }\href {https://doi.org/10.1103/PhysRevLett.125.266804}
  {\bibfield  {journal} {\bibinfo  {journal} {Phys. Rev. Lett.}\ }\textbf
  {\bibinfo {volume} {125}},\ \bibinfo {pages} {266804} (\bibinfo {year}
  {2020})}\BibitemShut {NoStop}%
\bibitem [{\citenamefont {Wang}\ \emph
  {et~al.}(2020{\natexlab{a}})\citenamefont {Wang}, \citenamefont {Lin},
  \citenamefont {Jiang}, \citenamefont {Guo},\ and\ \citenamefont
  {Jiang}}]{PhysRevLett.125.146401}%
  \BibitemOpen
  \bibfield  {author} {\bibinfo {author} {\bibfnamefont {H.-X.}\ \bibnamefont
  {Wang}}, \bibinfo {author} {\bibfnamefont {Z.-K.}\ \bibnamefont {Lin}},
  \bibinfo {author} {\bibfnamefont {B.}~\bibnamefont {Jiang}}, \bibinfo
  {author} {\bibfnamefont {G.-Y.}\ \bibnamefont {Guo}},\ and\ \bibinfo {author}
  {\bibfnamefont {J.-H.}\ \bibnamefont {Jiang}},\ }\bibfield  {title} {\bibinfo
  {title} {Higher-order {W}eyl semimetals},\ }\href
  {https://doi.org/10.1103/PhysRevLett.125.146401} {\bibfield  {journal}
  {\bibinfo  {journal} {Phys. Rev. Lett.}\ }\textbf {\bibinfo {volume} {125}},\
  \bibinfo {pages} {146401} (\bibinfo {year} {2020}{\natexlab{a}})}\BibitemShut
  {NoStop}%
\bibitem [{\citenamefont {Du}\ \emph {et~al.}(2022)\citenamefont {Du},
  \citenamefont {Chen}, \citenamefont {Wang},\ and\ \citenamefont
  {Xu}}]{PhysRevB.105.L081102}%
  \BibitemOpen
  \bibfield  {author} {\bibinfo {author} {\bibfnamefont {X.-L.}\ \bibnamefont
  {Du}}, \bibinfo {author} {\bibfnamefont {R.}~\bibnamefont {Chen}}, \bibinfo
  {author} {\bibfnamefont {R.}~\bibnamefont {Wang}},\ and\ \bibinfo {author}
  {\bibfnamefont {D.-H.}\ \bibnamefont {Xu}},\ }\bibfield  {title} {\bibinfo
  {title} {Weyl nodes with higher-order topology in an optically driven
  nodal-line semimetal},\ }\href {https://doi.org/10.1103/PhysRevB.105.L081102}
  {\bibfield  {journal} {\bibinfo  {journal} {Phys. Rev. B}\ }\textbf {\bibinfo
  {volume} {105}},\ \bibinfo {pages} {L081102} (\bibinfo {year}
  {2022})}\BibitemShut {NoStop}%
\bibitem [{\citenamefont {Wei}\ \emph {et~al.}(2021)\citenamefont {Wei},
  \citenamefont {Zhang}, \citenamefont {Deng}, \citenamefont {Lu},
  \citenamefont {Huang}, \citenamefont {Yan}, \citenamefont {Chen},
  \citenamefont {Liu},\ and\ \citenamefont {Jia}}]{Wei_2021}%
  \BibitemOpen
  \bibfield  {author} {\bibinfo {author} {\bibfnamefont {Q.}~\bibnamefont
  {Wei}}, \bibinfo {author} {\bibfnamefont {X.}~\bibnamefont {Zhang}}, \bibinfo
  {author} {\bibfnamefont {W.}~\bibnamefont {Deng}}, \bibinfo {author}
  {\bibfnamefont {J.}~\bibnamefont {Lu}}, \bibinfo {author} {\bibfnamefont
  {X.}~\bibnamefont {Huang}}, \bibinfo {author} {\bibfnamefont
  {M.}~\bibnamefont {Yan}}, \bibinfo {author} {\bibfnamefont {G.}~\bibnamefont
  {Chen}}, \bibinfo {author} {\bibfnamefont {Z.}~\bibnamefont {Liu}},\ and\
  \bibinfo {author} {\bibfnamefont {S.}~\bibnamefont {Jia}},\ }\bibfield
  {title} {\bibinfo {title} {Higher-order topological semimetal in acoustic
  crystals},\ }\href {https://doi.org/10.1038/s41563-021-00933-4} {\bibfield
  {journal} {\bibinfo  {journal} {Nature Materials}\ }\textbf {\bibinfo
  {volume} {20}},\ \bibinfo {pages} {812} (\bibinfo {year} {2021})}\BibitemShut
  {NoStop}%
\bibitem [{\citenamefont {Chen}\ \emph {et~al.}(2022)\citenamefont {Chen},
  \citenamefont {Zeng}, \citenamefont {Chen}, \citenamefont {Zhao},
  \citenamefont {Sheng},\ and\ \citenamefont {Yang}}]{PhysRevLett.128.026405}%
  \BibitemOpen
  \bibfield  {author} {\bibinfo {author} {\bibfnamefont {C.}~\bibnamefont
  {Chen}}, \bibinfo {author} {\bibfnamefont {X.-T.}\ \bibnamefont {Zeng}},
  \bibinfo {author} {\bibfnamefont {Z.}~\bibnamefont {Chen}}, \bibinfo {author}
  {\bibfnamefont {Y.~X.}\ \bibnamefont {Zhao}}, \bibinfo {author}
  {\bibfnamefont {X.-L.}\ \bibnamefont {Sheng}},\ and\ \bibinfo {author}
  {\bibfnamefont {S.~A.}\ \bibnamefont {Yang}},\ }\bibfield  {title} {\bibinfo
  {title} {Second-order real nodal-line semimetal in three-dimensional
  graphdiyne},\ }\href {https://doi.org/10.1103/PhysRevLett.128.026405}
  {\bibfield  {journal} {\bibinfo  {journal} {Phys. Rev. Lett.}\ }\textbf
  {\bibinfo {volume} {128}},\ \bibinfo {pages} {026405} (\bibinfo {year}
  {2022})}\BibitemShut {NoStop}%
\bibitem [{\citenamefont {Kondo}\ and\ \citenamefont
  {Akagi}(2021)}]{PhysRevLett.127.177201}%
  \BibitemOpen
  \bibfield  {author} {\bibinfo {author} {\bibfnamefont {H.}~\bibnamefont
  {Kondo}}\ and\ \bibinfo {author} {\bibfnamefont {Y.}~\bibnamefont {Akagi}},\
  }\bibfield  {title} {\bibinfo {title} {Dirac surface states in magnonic
  analogs of topological crystalline insulators},\ }\href
  {https://doi.org/10.1103/PhysRevLett.127.177201} {\bibfield  {journal}
  {\bibinfo  {journal} {Phys. Rev. Lett.}\ }\textbf {\bibinfo {volume} {127}},\
  \bibinfo {pages} {177201} (\bibinfo {year} {2021})}\BibitemShut {NoStop}%
\bibitem [{\citenamefont {Liu}\ \emph {et~al.}(2021)\citenamefont {Liu},
  \citenamefont {He}, \citenamefont {Yang},\ and\ \citenamefont
  {Nori}}]{PhysRevLett.127.196801}%
  \BibitemOpen
  \bibfield  {author} {\bibinfo {author} {\bibfnamefont {T.}~\bibnamefont
  {Liu}}, \bibinfo {author} {\bibfnamefont {J.~J.}\ \bibnamefont {He}},
  \bibinfo {author} {\bibfnamefont {Z.}~\bibnamefont {Yang}},\ and\ \bibinfo
  {author} {\bibfnamefont {F.}~\bibnamefont {Nori}},\ }\bibfield  {title}
  {\bibinfo {title} {Higher-order {W}eyl-exceptional-ring semimetals},\ }\href
  {https://doi.org/10.1103/PhysRevLett.127.196801} {\bibfield  {journal}
  {\bibinfo  {journal} {Phys. Rev. Lett.}\ }\textbf {\bibinfo {volume} {127}},\
  \bibinfo {pages} {196801} (\bibinfo {year} {2021})}\BibitemShut {NoStop}%
\bibitem [{\citenamefont {Kruthoff}\ \emph {et~al.}(2017)\citenamefont
  {Kruthoff}, \citenamefont {de~Boer}, \citenamefont {van Wezel}, \citenamefont
  {Kane},\ and\ \citenamefont {Slager}}]{PhysRevX.7.041069}%
  \BibitemOpen
  \bibfield  {author} {\bibinfo {author} {\bibfnamefont {J.}~\bibnamefont
  {Kruthoff}}, \bibinfo {author} {\bibfnamefont {J.}~\bibnamefont {de~Boer}},
  \bibinfo {author} {\bibfnamefont {J.}~\bibnamefont {van Wezel}}, \bibinfo
  {author} {\bibfnamefont {C.~L.}\ \bibnamefont {Kane}},\ and\ \bibinfo
  {author} {\bibfnamefont {R.-J.}\ \bibnamefont {Slager}},\ }\bibfield  {title}
  {\bibinfo {title} {Topological classification of crystalline insulators
  through band structure combinatorics},\ }\href
  {https://doi.org/10.1103/PhysRevX.7.041069} {\bibfield  {journal} {\bibinfo
  {journal} {Phys. Rev. X}\ }\textbf {\bibinfo {volume} {7}},\ \bibinfo {pages}
  {041069} (\bibinfo {year} {2017})}\BibitemShut {NoStop}%
\bibitem [{\citenamefont {Geier}\ \emph {et~al.}(2018)\citenamefont {Geier},
  \citenamefont {Trifunovic}, \citenamefont {Hoskam},\ and\ \citenamefont
  {Brouwer}}]{PhysRevB.97.205135}%
  \BibitemOpen
  \bibfield  {author} {\bibinfo {author} {\bibfnamefont {M.}~\bibnamefont
  {Geier}}, \bibinfo {author} {\bibfnamefont {L.}~\bibnamefont {Trifunovic}},
  \bibinfo {author} {\bibfnamefont {M.}~\bibnamefont {Hoskam}},\ and\ \bibinfo
  {author} {\bibfnamefont {P.~W.}\ \bibnamefont {Brouwer}},\ }\bibfield
  {title} {\bibinfo {title} {Second-order topological insulators and
  superconductors with an order-two crystalline symmetry},\ }\href
  {https://doi.org/10.1103/PhysRevB.97.205135} {\bibfield  {journal} {\bibinfo
  {journal} {Phys. Rev. B}\ }\textbf {\bibinfo {volume} {97}},\ \bibinfo
  {pages} {205135} (\bibinfo {year} {2018})}\BibitemShut {NoStop}%
\bibitem [{\citenamefont {Huang}\ \emph {et~al.}(2024)\citenamefont {Huang},
  \citenamefont {Park},\ and\ \citenamefont {Hsu}}]{shengjiehuang}%
  \BibitemOpen
  \bibfield  {author} {\bibinfo {author} {\bibfnamefont {S.-J.}\ \bibnamefont
  {Huang}}, \bibinfo {author} {\bibfnamefont {K.}~\bibnamefont {Park}},\ and\
  \bibinfo {author} {\bibfnamefont {Y.-T.}\ \bibnamefont {Hsu}},\ }\bibfield
  {title} {\bibinfo {title} {Hybrid-order topological superconductivity in a
  topological metal 1{T}'-{M}o{T}e$_2$},\ }\href
  {https://doi.org/10.1038/s41535-024-00633-7} {\bibfield  {journal} {\bibinfo
  {journal} {npj Quantum Materials}\ }\textbf {\bibinfo {volume} {9}},\
  \bibinfo {pages} {21} (\bibinfo {year} {2024})}\BibitemShut {NoStop}%
\bibitem [{\citenamefont {Yang}\ \emph {et~al.}(2021)\citenamefont {Yang},
  \citenamefont {Lu}, \citenamefont {Yan}, \citenamefont {Huang}, \citenamefont
  {Deng},\ and\ \citenamefont {Liu}}]{PhysRevLett.126.156801}%
  \BibitemOpen
  \bibfield  {author} {\bibinfo {author} {\bibfnamefont {Y.}~\bibnamefont
  {Yang}}, \bibinfo {author} {\bibfnamefont {J.}~\bibnamefont {Lu}}, \bibinfo
  {author} {\bibfnamefont {M.}~\bibnamefont {Yan}}, \bibinfo {author}
  {\bibfnamefont {X.}~\bibnamefont {Huang}}, \bibinfo {author} {\bibfnamefont
  {W.}~\bibnamefont {Deng}},\ and\ \bibinfo {author} {\bibfnamefont
  {Z.}~\bibnamefont {Liu}},\ }\bibfield  {title} {\bibinfo {title}
  {Hybrid-order topological insulators in a phononic crystal},\ }\href
  {https://doi.org/10.1103/PhysRevLett.126.156801} {\bibfield  {journal}
  {\bibinfo  {journal} {Phys. Rev. Lett.}\ }\textbf {\bibinfo {volume} {126}},\
  \bibinfo {pages} {156801} (\bibinfo {year} {2021})}\BibitemShut {NoStop}%
\bibitem [{\citenamefont {Zhang}\ \emph {et~al.}(2020)\citenamefont {Zhang},
  \citenamefont {Lin}, \citenamefont {Wang}, \citenamefont {Xiong},
  \citenamefont {Tian}, \citenamefont {Lu}, \citenamefont {Chen},\ and\
  \citenamefont {Jiang}}]{xiujuanzhang}%
  \BibitemOpen
  \bibfield  {author} {\bibinfo {author} {\bibfnamefont {X.}~\bibnamefont
  {Zhang}}, \bibinfo {author} {\bibfnamefont {Z.-K.}\ \bibnamefont {Lin}},
  \bibinfo {author} {\bibfnamefont {H.-X.}\ \bibnamefont {Wang}}, \bibinfo
  {author} {\bibfnamefont {Z.}~\bibnamefont {Xiong}}, \bibinfo {author}
  {\bibfnamefont {Y.}~\bibnamefont {Tian}}, \bibinfo {author} {\bibfnamefont
  {M.-H.}\ \bibnamefont {Lu}}, \bibinfo {author} {\bibfnamefont {Y.-F.}\
  \bibnamefont {Chen}},\ and\ \bibinfo {author} {\bibfnamefont {J.-H.}\
  \bibnamefont {Jiang}},\ }\bibfield  {title} {\bibinfo {title}
  {Symmetry-protected hierarchy of anomalous multipole topological band gaps in
  nonsymmorphic metacrystals},\ }\href
  {https://doi.org/10.1038/s41467-019-13861-4} {\bibfield  {journal} {\bibinfo
  {journal} {Nature Communications}\ }\textbf {\bibinfo {volume} {11}},\
  \bibinfo {pages} {65} (\bibinfo {year} {2020})}\BibitemShut {NoStop}%
\bibitem [{\citenamefont {Ahn}\ \emph {et~al.}(2019)\citenamefont {Ahn},
  \citenamefont {Park},\ and\ \citenamefont {Yang}}]{PhysRevX.9.021013}%
  \BibitemOpen
  \bibfield  {author} {\bibinfo {author} {\bibfnamefont {J.}~\bibnamefont
  {Ahn}}, \bibinfo {author} {\bibfnamefont {S.}~\bibnamefont {Park}},\ and\
  \bibinfo {author} {\bibfnamefont {B.-J.}\ \bibnamefont {Yang}},\ }\bibfield
  {title} {\bibinfo {title} {Failure of {N}ielsen-{N}inomiya theorem and
  fragile topology in two-dimensional systems with space-time inversion
  symmetry: Application to twisted bilayer graphene at magic angle},\ }\href
  {https://doi.org/10.1103/PhysRevX.9.021013} {\bibfield  {journal} {\bibinfo
  {journal} {Phys. Rev. X}\ }\textbf {\bibinfo {volume} {9}},\ \bibinfo {pages}
  {021013} (\bibinfo {year} {2019})}\BibitemShut {NoStop}%
\bibitem [{\citenamefont {Sheng}\ \emph {et~al.}(2019)\citenamefont {Sheng},
  \citenamefont {Chen}, \citenamefont {Liu}, \citenamefont {Chen},
  \citenamefont {Yu}, \citenamefont {Zhao},\ and\ \citenamefont
  {Yang}}]{PhysRevLett.123.256402}%
  \BibitemOpen
  \bibfield  {author} {\bibinfo {author} {\bibfnamefont {X.-L.}\ \bibnamefont
  {Sheng}}, \bibinfo {author} {\bibfnamefont {C.}~\bibnamefont {Chen}},
  \bibinfo {author} {\bibfnamefont {H.}~\bibnamefont {Liu}}, \bibinfo {author}
  {\bibfnamefont {Z.}~\bibnamefont {Chen}}, \bibinfo {author} {\bibfnamefont
  {Z.-M.}\ \bibnamefont {Yu}}, \bibinfo {author} {\bibfnamefont {Y.~X.}\
  \bibnamefont {Zhao}},\ and\ \bibinfo {author} {\bibfnamefont {S.~A.}\
  \bibnamefont {Yang}},\ }\bibfield  {title} {\bibinfo {title} {Two-dimensional
  second-order topological insulator in graphdiyne},\ }\href
  {https://doi.org/10.1103/PhysRevLett.123.256402} {\bibfield  {journal}
  {\bibinfo  {journal} {Phys. Rev. Lett.}\ }\textbf {\bibinfo {volume} {123}},\
  \bibinfo {pages} {256402} (\bibinfo {year} {2019})}\BibitemShut {NoStop}%
\bibitem [{\citenamefont {Wang}\ \emph
  {et~al.}(2024{\natexlab{a}})\citenamefont {Wang}, \citenamefont {Cui},
  \citenamefont {Zhang}, \citenamefont {Wang}, \citenamefont {Yu},
  \citenamefont {Liu},\ and\ \citenamefont
  {Yao}}]{wang2024mirrorrealcherninsulator}%
  \BibitemOpen
  \bibfield  {author} {\bibinfo {author} {\bibfnamefont {Y.}~\bibnamefont
  {Wang}}, \bibinfo {author} {\bibfnamefont {C.}~\bibnamefont {Cui}}, \bibinfo
  {author} {\bibfnamefont {R.-W.}\ \bibnamefont {Zhang}}, \bibinfo {author}
  {\bibfnamefont {X.}~\bibnamefont {Wang}}, \bibinfo {author} {\bibfnamefont
  {Z.-M.}\ \bibnamefont {Yu}}, \bibinfo {author} {\bibfnamefont {G.-B.}\
  \bibnamefont {Liu}},\ and\ \bibinfo {author} {\bibfnamefont {Y.}~\bibnamefont
  {Yao}},\ }\href {https://arxiv.org/abs/2403.01145} {\bibinfo {title} {Mirror
  real {C}hern insulator in two and three dimensions}} (\bibinfo {year}
  {2024}{\natexlab{a}}),\ \Eprint {https://arxiv.org/abs/2403.01145}
  {arXiv:2403.01145 [cond-mat.mtrl-sci]} \BibitemShut {NoStop}%
\bibitem [{\citenamefont {Ozawa}\ \emph {et~al.}(2019)\citenamefont {Ozawa},
  \citenamefont {Price}, \citenamefont {Amo}, \citenamefont {Goldman},
  \citenamefont {Hafezi}, \citenamefont {Lu}, \citenamefont {Rechtsman},
  \citenamefont {Schuster}, \citenamefont {Simon}, \citenamefont {Zilberberg},\
  and\ \citenamefont {Carusotto}}]{RevModPhys.91.015006}%
  \BibitemOpen
  \bibfield  {author} {\bibinfo {author} {\bibfnamefont {T.}~\bibnamefont
  {Ozawa}}, \bibinfo {author} {\bibfnamefont {H.~M.}\ \bibnamefont {Price}},
  \bibinfo {author} {\bibfnamefont {A.}~\bibnamefont {Amo}}, \bibinfo {author}
  {\bibfnamefont {N.}~\bibnamefont {Goldman}}, \bibinfo {author} {\bibfnamefont
  {M.}~\bibnamefont {Hafezi}}, \bibinfo {author} {\bibfnamefont
  {L.}~\bibnamefont {Lu}}, \bibinfo {author} {\bibfnamefont {M.~C.}\
  \bibnamefont {Rechtsman}}, \bibinfo {author} {\bibfnamefont {D.}~\bibnamefont
  {Schuster}}, \bibinfo {author} {\bibfnamefont {J.}~\bibnamefont {Simon}},
  \bibinfo {author} {\bibfnamefont {O.}~\bibnamefont {Zilberberg}},\ and\
  \bibinfo {author} {\bibfnamefont {I.}~\bibnamefont {Carusotto}},\ }\bibfield
  {title} {\bibinfo {title} {Topological photonics},\ }\href
  {https://doi.org/10.1103/RevModPhys.91.015006} {\bibfield  {journal}
  {\bibinfo  {journal} {Rev. Mod. Phys.}\ }\textbf {\bibinfo {volume} {91}},\
  \bibinfo {pages} {015006} (\bibinfo {year} {2019})}\BibitemShut {NoStop}%
\bibitem [{\citenamefont {Yang}\ \emph {et~al.}(2015)\citenamefont {Yang},
  \citenamefont {Gao}, \citenamefont {Shi}, \citenamefont {Lin}, \citenamefont
  {Gao}, \citenamefont {Chong},\ and\ \citenamefont
  {Zhang}}]{PhysRevLett.114.114301}%
  \BibitemOpen
  \bibfield  {author} {\bibinfo {author} {\bibfnamefont {Z.}~\bibnamefont
  {Yang}}, \bibinfo {author} {\bibfnamefont {F.}~\bibnamefont {Gao}}, \bibinfo
  {author} {\bibfnamefont {X.}~\bibnamefont {Shi}}, \bibinfo {author}
  {\bibfnamefont {X.}~\bibnamefont {Lin}}, \bibinfo {author} {\bibfnamefont
  {Z.}~\bibnamefont {Gao}}, \bibinfo {author} {\bibfnamefont {Y.}~\bibnamefont
  {Chong}},\ and\ \bibinfo {author} {\bibfnamefont {B.}~\bibnamefont {Zhang}},\
  }\bibfield  {title} {\bibinfo {title} {Topological acoustics},\ }\href
  {https://doi.org/10.1103/PhysRevLett.114.114301} {\bibfield  {journal}
  {\bibinfo  {journal} {Phys. Rev. Lett.}\ }\textbf {\bibinfo {volume} {114}},\
  \bibinfo {pages} {114301} (\bibinfo {year} {2015})}\BibitemShut {NoStop}%
\bibitem [{\citenamefont {Wang}\ \emph
  {et~al.}(2024{\natexlab{b}})\citenamefont {Wang}, \citenamefont {Liu},
  \citenamefont {Ding},\ and\ \citenamefont {He}}]{PhysRevA.109.053314}%
  \BibitemOpen
  \bibfield  {author} {\bibinfo {author} {\bibfnamefont {J.-T.}\ \bibnamefont
  {Wang}}, \bibinfo {author} {\bibfnamefont {J.-X.}\ \bibnamefont {Liu}},
  \bibinfo {author} {\bibfnamefont {H.-T.}\ \bibnamefont {Ding}},\ and\
  \bibinfo {author} {\bibfnamefont {P.}~\bibnamefont {He}},\ }\bibfield
  {title} {\bibinfo {title} {Proposal for implementing {S}tiefel-{W}hitney
  insulators in an optical {R}aman lattice},\ }\href
  {https://doi.org/10.1103/PhysRevA.109.053314} {\bibfield  {journal} {\bibinfo
   {journal} {Phys. Rev. A}\ }\textbf {\bibinfo {volume} {109}},\ \bibinfo
  {pages} {053314} (\bibinfo {year} {2024}{\natexlab{b}})}\BibitemShut
  {NoStop}%
\bibitem [{\citenamefont {Takeichi}\ \emph {et~al.}(2023)\citenamefont
  {Takeichi}, \citenamefont {Furuta},\ and\ \citenamefont
  {Murakami}}]{PhysRevB.107.085139}%
  \BibitemOpen
  \bibfield  {author} {\bibinfo {author} {\bibfnamefont {M.}~\bibnamefont
  {Takeichi}}, \bibinfo {author} {\bibfnamefont {R.}~\bibnamefont {Furuta}},\
  and\ \bibinfo {author} {\bibfnamefont {S.}~\bibnamefont {Murakami}},\
  }\bibfield  {title} {\bibinfo {title} {Morse theory study on the evolution of
  nodal lines in $\mathcal{PT}$-symmetric nodal-line semimetals},\ }\href
  {https://doi.org/10.1103/PhysRevB.107.085139} {\bibfield  {journal} {\bibinfo
   {journal} {Phys. Rev. B}\ }\textbf {\bibinfo {volume} {107}},\ \bibinfo
  {pages} {085139} (\bibinfo {year} {2023})}\BibitemShut {NoStop}%
\bibitem [{\citenamefont {Song}\ \emph {et~al.}(2018)\citenamefont {Song},
  \citenamefont {Zhang},\ and\ \citenamefont {Fang}}]{PhysRevX.8.031069}%
  \BibitemOpen
  \bibfield  {author} {\bibinfo {author} {\bibfnamefont {Z.}~\bibnamefont
  {Song}}, \bibinfo {author} {\bibfnamefont {T.}~\bibnamefont {Zhang}},\ and\
  \bibinfo {author} {\bibfnamefont {C.}~\bibnamefont {Fang}},\ }\bibfield
  {title} {\bibinfo {title} {Diagnosis for nonmagnetic topological semimetals
  in the absence of spin-orbital coupling},\ }\href
  {https://doi.org/10.1103/PhysRevX.8.031069} {\bibfield  {journal} {\bibinfo
  {journal} {Phys. Rev. X}\ }\textbf {\bibinfo {volume} {8}},\ \bibinfo {pages}
  {031069} (\bibinfo {year} {2018})}\BibitemShut {NoStop}%
\bibitem [{\citenamefont {Pan}\ and\ \citenamefont
  {Huang}(2022)}]{PhysRevB.106.L201406}%
  \BibitemOpen
  \bibfield  {author} {\bibinfo {author} {\bibfnamefont {M.}~\bibnamefont
  {Pan}}\ and\ \bibinfo {author} {\bibfnamefont {H.}~\bibnamefont {Huang}},\
  }\bibfield  {title} {\bibinfo {title} {Phononic stiefel-whitney topology with
  corner vibrational modes in two-dimensional xenes and ligand-functionalized
  derivatives},\ }\href {https://doi.org/10.1103/PhysRevB.106.L201406}
  {\bibfield  {journal} {\bibinfo  {journal} {Phys. Rev. B}\ }\textbf {\bibinfo
  {volume} {106}},\ \bibinfo {pages} {L201406} (\bibinfo {year}
  {2022})}\BibitemShut {NoStop}%
\bibitem [{\citenamefont {Takahashi}\ and\ \citenamefont
  {Ozawa}(2024)}]{PhysRevResearch.6.033192}%
  \BibitemOpen
  \bibfield  {author} {\bibinfo {author} {\bibfnamefont {R.}~\bibnamefont
  {Takahashi}}\ and\ \bibinfo {author} {\bibfnamefont {T.}~\bibnamefont
  {Ozawa}},\ }\bibfield  {title} {\bibinfo {title} {Bulk-entanglement spectrum
  correspondence in $pt$- and $pc$-symmetric topological insulators and
  superconductors},\ }\href {https://doi.org/10.1103/PhysRevResearch.6.033192}
  {\bibfield  {journal} {\bibinfo  {journal} {Phys. Rev. Res.}\ }\textbf
  {\bibinfo {volume} {6}},\ \bibinfo {pages} {033192} (\bibinfo {year}
  {2024})}\BibitemShut {NoStop}%
\bibitem [{\citenamefont {Takahashi}\ and\ \citenamefont
  {Ozawa}(2023)}]{PhysRevB.108.075129}%
  \BibitemOpen
  \bibfield  {author} {\bibinfo {author} {\bibfnamefont {R.}~\bibnamefont
  {Takahashi}}\ and\ \bibinfo {author} {\bibfnamefont {T.}~\bibnamefont
  {Ozawa}},\ }\bibfield  {title} {\bibinfo {title} {Bulk-edge correspondence of
  stiefel-whitney and euler insulators through the entanglement spectrum and
  cutting procedure},\ }\href {https://doi.org/10.1103/PhysRevB.108.075129}
  {\bibfield  {journal} {\bibinfo  {journal} {Phys. Rev. B}\ }\textbf {\bibinfo
  {volume} {108}},\ \bibinfo {pages} {075129} (\bibinfo {year}
  {2023})}\BibitemShut {NoStop}%
\bibitem [{\citenamefont {Wu}\ \emph {et~al.}(2019)\citenamefont {Wu},
  \citenamefont {Soluyanov},\ and\ \citenamefont {Bzdušek}}]{Wu_2019}%
  \BibitemOpen
  \bibfield  {author} {\bibinfo {author} {\bibfnamefont {Q.}~\bibnamefont
  {Wu}}, \bibinfo {author} {\bibfnamefont {A.~A.}\ \bibnamefont {Soluyanov}},\
  and\ \bibinfo {author} {\bibfnamefont {T.}~\bibnamefont {Bzdušek}},\
  }\bibfield  {title} {\bibinfo {title} {Non-{A}belian band topology in
  noninteracting metals},\ }\href {https://doi.org/10.1126/science.aau8740}
  {\bibfield  {journal} {\bibinfo  {journal} {Science}\ }\textbf {\bibinfo
  {volume} {365}},\ \bibinfo {pages} {1273–1277} (\bibinfo {year}
  {2019})}\BibitemShut {NoStop}%
\bibitem [{\citenamefont {Guo}\ \emph {et~al.}(2021)\citenamefont {Guo},
  \citenamefont {Jiang}, \citenamefont {Zhang}, \citenamefont {Zhang},
  \citenamefont {Zhang}, \citenamefont {Yang}, \citenamefont {Zhang},\ and\
  \citenamefont {Chan}}]{Guo_2021}%
  \BibitemOpen
  \bibfield  {author} {\bibinfo {author} {\bibfnamefont {Q.}~\bibnamefont
  {Guo}}, \bibinfo {author} {\bibfnamefont {T.}~\bibnamefont {Jiang}}, \bibinfo
  {author} {\bibfnamefont {R.-Y.}\ \bibnamefont {Zhang}}, \bibinfo {author}
  {\bibfnamefont {L.}~\bibnamefont {Zhang}}, \bibinfo {author} {\bibfnamefont
  {Z.-Q.}\ \bibnamefont {Zhang}}, \bibinfo {author} {\bibfnamefont
  {B.}~\bibnamefont {Yang}}, \bibinfo {author} {\bibfnamefont {S.}~\bibnamefont
  {Zhang}},\ and\ \bibinfo {author} {\bibfnamefont {C.~T.}\ \bibnamefont
  {Chan}},\ }\bibfield  {title} {\bibinfo {title} {Experimental observation of
  non-{A}belian topological charges and edge states},\ }\href
  {https://doi.org/10.1038/s41586-021-03521-3} {\bibfield  {journal} {\bibinfo
  {journal} {Nature}\ }\textbf {\bibinfo {volume} {594}},\ \bibinfo {pages}
  {195–200} (\bibinfo {year} {2021})}\BibitemShut {NoStop}%
\bibitem [{\citenamefont {Sun}\ \emph {et~al.}(2024)\citenamefont {Sun},
  \citenamefont {Wang}, \citenamefont {He},\ and\ \citenamefont
  {Chen}}]{PhysRevLett.132.216602}%
  \BibitemOpen
  \bibfield  {author} {\bibinfo {author} {\bibfnamefont {X.-C.}\ \bibnamefont
  {Sun}}, \bibinfo {author} {\bibfnamefont {J.-B.}\ \bibnamefont {Wang}},
  \bibinfo {author} {\bibfnamefont {C.}~\bibnamefont {He}},\ and\ \bibinfo
  {author} {\bibfnamefont {Y.-F.}\ \bibnamefont {Chen}},\ }\bibfield  {title}
  {\bibinfo {title} {Non-abelian topological phases and their quotient
  relations in acoustic systems},\ }\href
  {https://doi.org/10.1103/PhysRevLett.132.216602} {\bibfield  {journal}
  {\bibinfo  {journal} {Phys. Rev. Lett.}\ }\textbf {\bibinfo {volume} {132}},\
  \bibinfo {pages} {216602} (\bibinfo {year} {2024})}\BibitemShut {NoStop}%
\bibitem [{\citenamefont {Jiang}\ \emph {et~al.}(2021)\citenamefont {Jiang},
  \citenamefont {Bouhon}, \citenamefont {Lin}, \citenamefont {Zhou},
  \citenamefont {Hou}, \citenamefont {Li}, \citenamefont {Slager},\ and\
  \citenamefont {Jiang}}]{Jiang_2021}%
  \BibitemOpen
  \bibfield  {author} {\bibinfo {author} {\bibfnamefont {B.}~\bibnamefont
  {Jiang}}, \bibinfo {author} {\bibfnamefont {A.}~\bibnamefont {Bouhon}},
  \bibinfo {author} {\bibfnamefont {Z.-K.}\ \bibnamefont {Lin}}, \bibinfo
  {author} {\bibfnamefont {X.}~\bibnamefont {Zhou}}, \bibinfo {author}
  {\bibfnamefont {B.}~\bibnamefont {Hou}}, \bibinfo {author} {\bibfnamefont
  {F.}~\bibnamefont {Li}}, \bibinfo {author} {\bibfnamefont {R.-J.}\
  \bibnamefont {Slager}},\ and\ \bibinfo {author} {\bibfnamefont {J.-H.}\
  \bibnamefont {Jiang}},\ }\bibfield  {title} {\bibinfo {title} {Experimental
  observation of non-abelian topological acoustic semimetals and their phase
  transitions},\ }\href {https://doi.org/10.1038/s41567-021-01340-x} {\bibfield
   {journal} {\bibinfo  {journal} {Nature Physics}\ }\textbf {\bibinfo {volume}
  {17}},\ \bibinfo {pages} {1239–1246} (\bibinfo {year} {2021})}\BibitemShut
  {NoStop}%
\bibitem [{\citenamefont {Bouhon}\ \emph {et~al.}(2020)\citenamefont {Bouhon},
  \citenamefont {Wu}, \citenamefont {Slager}, \citenamefont {Weng},
  \citenamefont {Yazyev},\ and\ \citenamefont {Bzdušek}}]{Bouhon_2020}%
  \BibitemOpen
  \bibfield  {author} {\bibinfo {author} {\bibfnamefont {A.}~\bibnamefont
  {Bouhon}}, \bibinfo {author} {\bibfnamefont {Q.}~\bibnamefont {Wu}}, \bibinfo
  {author} {\bibfnamefont {R.-J.}\ \bibnamefont {Slager}}, \bibinfo {author}
  {\bibfnamefont {H.}~\bibnamefont {Weng}}, \bibinfo {author} {\bibfnamefont
  {O.~V.}\ \bibnamefont {Yazyev}},\ and\ \bibinfo {author} {\bibfnamefont
  {T.}~\bibnamefont {Bzdušek}},\ }\bibfield  {title} {\bibinfo {title}
  {Non-abelian reciprocal braiding of {W}eyl points and its manifestation in
  {Z}r{T}e},\ }\href {https://doi.org/10.1038/s41567-020-0967-9} {\bibfield
  {journal} {\bibinfo  {journal} {Nature Physics}\ }\textbf {\bibinfo {volume}
  {16}},\ \bibinfo {pages} {1137–1143} (\bibinfo {year} {2020})}\BibitemShut
  {NoStop}%
\bibitem [{\citenamefont {Peng}\ \emph {et~al.}(2022)\citenamefont {Peng},
  \citenamefont {Bouhon}, \citenamefont {Monserrat},\ and\ \citenamefont
  {Slager}}]{pengbo}%
  \BibitemOpen
  \bibfield  {author} {\bibinfo {author} {\bibfnamefont {B.}~\bibnamefont
  {Peng}}, \bibinfo {author} {\bibfnamefont {A.}~\bibnamefont {Bouhon}},
  \bibinfo {author} {\bibfnamefont {B.}~\bibnamefont {Monserrat}},\ and\
  \bibinfo {author} {\bibfnamefont {R.-J.}\ \bibnamefont {Slager}},\ }\bibfield
   {title} {\bibinfo {title} {Phonons as a platform for non-abelian braiding
  and its manifestation in layered silicates},\ }\href
  {https://doi.org/10.1038/s41467-022-28046-9} {\bibfield  {journal} {\bibinfo
  {journal} {Nature Communications}\ }\textbf {\bibinfo {volume} {13}},\
  \bibinfo {pages} {423} (\bibinfo {year} {2022})}\BibitemShut {NoStop}%
\bibitem [{\citenamefont {Jiang}\ \emph {et~al.}(2024)\citenamefont {Jiang},
  \citenamefont {Bouhon}, \citenamefont {Wu}, \citenamefont {Kong},
  \citenamefont {Lin}, \citenamefont {Slager},\ and\ \citenamefont
  {Jiang}}]{jiang2022experimentalobservationmeronictopological}%
  \BibitemOpen
  \bibfield  {author} {\bibinfo {author} {\bibfnamefont {B.}~\bibnamefont
  {Jiang}}, \bibinfo {author} {\bibfnamefont {A.}~\bibnamefont {Bouhon}},
  \bibinfo {author} {\bibfnamefont {S.-Q.}\ \bibnamefont {Wu}}, \bibinfo
  {author} {\bibfnamefont {Z.-L.}\ \bibnamefont {Kong}}, \bibinfo {author}
  {\bibfnamefont {Z.-K.}\ \bibnamefont {Lin}}, \bibinfo {author} {\bibfnamefont
  {R.-J.}\ \bibnamefont {Slager}},\ and\ \bibinfo {author} {\bibfnamefont
  {J.-H.}\ \bibnamefont {Jiang}},\ }\bibfield  {title} {\bibinfo {title}
  {Experimental observation of meronic topological acoustic {E}uler
  insulators},\ }\href {https://doi.org/10.1016/j.scib.2024.04.009} {\bibfield
  {journal} {\bibinfo  {journal} {Science Bulletin}\ }\textbf {\bibinfo
  {volume} {69}},\ \bibinfo {pages} {1653} (\bibinfo {year}
  {2024})}\BibitemShut {NoStop}%
\bibitem [{\citenamefont {Wang}\ \emph
  {et~al.}(2020{\natexlab{b}})\citenamefont {Wang}, \citenamefont {Dai},
  \citenamefont {Shao}, \citenamefont {Yang},\ and\ \citenamefont
  {Zhao}}]{PhysRevLett.125.126403}%
  \BibitemOpen
  \bibfield  {author} {\bibinfo {author} {\bibfnamefont {K.}~\bibnamefont
  {Wang}}, \bibinfo {author} {\bibfnamefont {J.-X.}\ \bibnamefont {Dai}},
  \bibinfo {author} {\bibfnamefont {L.~B.}\ \bibnamefont {Shao}}, \bibinfo
  {author} {\bibfnamefont {S.~A.}\ \bibnamefont {Yang}},\ and\ \bibinfo
  {author} {\bibfnamefont {Y.~X.}\ \bibnamefont {Zhao}},\ }\bibfield  {title}
  {\bibinfo {title} {Boundary criticality of $\mathcal{PT}$-invariant topology
  and second-order nodal-line semimetals},\ }\href
  {https://doi.org/10.1103/PhysRevLett.125.126403} {\bibfield  {journal}
  {\bibinfo  {journal} {Phys. Rev. Lett.}\ }\textbf {\bibinfo {volume} {125}},\
  \bibinfo {pages} {126403} (\bibinfo {year} {2020}{\natexlab{b}})}\BibitemShut
  {NoStop}%
\bibitem [{\citenamefont {Zhao}\ and\ \citenamefont
  {Lu}(2017)}]{PhysRevLett.118.056401}%
  \BibitemOpen
  \bibfield  {author} {\bibinfo {author} {\bibfnamefont {Y.~X.}\ \bibnamefont
  {Zhao}}\ and\ \bibinfo {author} {\bibfnamefont {Y.}~\bibnamefont {Lu}},\
  }\bibfield  {title} {\bibinfo {title} {$pt$-symmetric real dirac fermions and
  semimetals},\ }\href {https://doi.org/10.1103/PhysRevLett.118.056401}
  {\bibfield  {journal} {\bibinfo  {journal} {Phys. Rev. Lett.}\ }\textbf
  {\bibinfo {volume} {118}},\ \bibinfo {pages} {056401} (\bibinfo {year}
  {2017})}\BibitemShut {NoStop}%
\bibitem [{\citenamefont {Ahn}\ \emph {et~al.}(2018)\citenamefont {Ahn},
  \citenamefont {Kim}, \citenamefont {Kim},\ and\ \citenamefont
  {Yang}}]{PhysRevLett.121.106403}%
  \BibitemOpen
  \bibfield  {author} {\bibinfo {author} {\bibfnamefont {J.}~\bibnamefont
  {Ahn}}, \bibinfo {author} {\bibfnamefont {D.}~\bibnamefont {Kim}}, \bibinfo
  {author} {\bibfnamefont {Y.}~\bibnamefont {Kim}},\ and\ \bibinfo {author}
  {\bibfnamefont {B.-J.}\ \bibnamefont {Yang}},\ }\bibfield  {title} {\bibinfo
  {title} {Band topology and linking structure of nodal line semimetals with
  ${Z}_{2}$ monopole charges},\ }\href
  {https://doi.org/10.1103/PhysRevLett.121.106403} {\bibfield  {journal}
  {\bibinfo  {journal} {Phys. Rev. Lett.}\ }\textbf {\bibinfo {volume} {121}},\
  \bibinfo {pages} {106403} (\bibinfo {year} {2018})}\BibitemShut {NoStop}%
\bibitem [{\citenamefont {Shiozaki}\ \emph {et~al.}(2017)\citenamefont
  {Shiozaki}, \citenamefont {Sato},\ and\ \citenamefont
  {Gomi}}]{PhysRevB.95.235425}%
  \BibitemOpen
  \bibfield  {author} {\bibinfo {author} {\bibfnamefont {K.}~\bibnamefont
  {Shiozaki}}, \bibinfo {author} {\bibfnamefont {M.}~\bibnamefont {Sato}},\
  and\ \bibinfo {author} {\bibfnamefont {K.}~\bibnamefont {Gomi}},\ }\bibfield
  {title} {\bibinfo {title} {Topological crystalline materials: General
  formulation, module structure, and wallpaper groups},\ }\href
  {https://doi.org/10.1103/PhysRevB.95.235425} {\bibfield  {journal} {\bibinfo
  {journal} {Phys. Rev. B}\ }\textbf {\bibinfo {volume} {95}},\ \bibinfo
  {pages} {235425} (\bibinfo {year} {2017})}\BibitemShut {NoStop}%
\bibitem [{\citenamefont {Yue}\ \emph {et~al.}(2024)\citenamefont {Yue},
  \citenamefont {Liu}, \citenamefont {Yang},\ and\ \citenamefont
  {Zhao}}]{PhysRevB.109.195116}%
  \BibitemOpen
  \bibfield  {author} {\bibinfo {author} {\bibfnamefont {S.~J.}\ \bibnamefont
  {Yue}}, \bibinfo {author} {\bibfnamefont {Q.}~\bibnamefont {Liu}}, \bibinfo
  {author} {\bibfnamefont {S.~A.}\ \bibnamefont {Yang}},\ and\ \bibinfo
  {author} {\bibfnamefont {Y.~X.}\ \bibnamefont {Zhao}},\ }\bibfield  {title}
  {\bibinfo {title} {Stability and noncentered ${PT}$ symmetry of real
  topological phases},\ }\href {https://doi.org/10.1103/PhysRevB.109.195116}
  {\bibfield  {journal} {\bibinfo  {journal} {Phys. Rev. B}\ }\textbf {\bibinfo
  {volume} {109}},\ \bibinfo {pages} {195116} (\bibinfo {year}
  {2024})}\BibitemShut {NoStop}%
\bibitem [{\citenamefont {Fang}\ \emph {et~al.}(2015)\citenamefont {Fang},
  \citenamefont {Chen}, \citenamefont {Kee},\ and\ \citenamefont
  {Fu}}]{PhysRevB.92.081201}%
  \BibitemOpen
  \bibfield  {author} {\bibinfo {author} {\bibfnamefont {C.}~\bibnamefont
  {Fang}}, \bibinfo {author} {\bibfnamefont {Y.}~\bibnamefont {Chen}}, \bibinfo
  {author} {\bibfnamefont {H.-Y.}\ \bibnamefont {Kee}},\ and\ \bibinfo {author}
  {\bibfnamefont {L.}~\bibnamefont {Fu}},\ }\bibfield  {title} {\bibinfo
  {title} {Topological nodal line semimetals with and without spin-orbital
  coupling},\ }\href {https://doi.org/10.1103/PhysRevB.92.081201} {\bibfield
  {journal} {\bibinfo  {journal} {Phys. Rev. B}\ }\textbf {\bibinfo {volume}
  {92}},\ \bibinfo {pages} {081201} (\bibinfo {year} {2015})}\BibitemShut
  {NoStop}%
\bibitem [{\citenamefont {Shao}\ \emph {et~al.}(2021)\citenamefont {Shao},
  \citenamefont {Liu}, \citenamefont {Xiao}, \citenamefont {Yang},\ and\
  \citenamefont {Zhao}}]{PhysRevLett.127.076401}%
  \BibitemOpen
  \bibfield  {author} {\bibinfo {author} {\bibfnamefont {L.~B.}\ \bibnamefont
  {Shao}}, \bibinfo {author} {\bibfnamefont {Q.}~\bibnamefont {Liu}}, \bibinfo
  {author} {\bibfnamefont {R.}~\bibnamefont {Xiao}}, \bibinfo {author}
  {\bibfnamefont {S.~A.}\ \bibnamefont {Yang}},\ and\ \bibinfo {author}
  {\bibfnamefont {Y.~X.}\ \bibnamefont {Zhao}},\ }\bibfield  {title} {\bibinfo
  {title} {Gauge-field extended $k\ifmmode\cdot\else\textperiodcentered\fi{}p$
  method and novel topological phases},\ }\href
  {https://doi.org/10.1103/PhysRevLett.127.076401} {\bibfield  {journal}
  {\bibinfo  {journal} {Phys. Rev. Lett.}\ }\textbf {\bibinfo {volume} {127}},\
  \bibinfo {pages} {076401} (\bibinfo {year} {2021})}\BibitemShut {NoStop}%
\bibitem [{\citenamefont {Zhao}\ \emph {et~al.}(2021)\citenamefont {Zhao},
  \citenamefont {Chen}, \citenamefont {Sheng},\ and\ \citenamefont
  {Yang}}]{PhysRevLett.126.196402}%
  \BibitemOpen
  \bibfield  {author} {\bibinfo {author} {\bibfnamefont {Y.~X.}\ \bibnamefont
  {Zhao}}, \bibinfo {author} {\bibfnamefont {C.}~\bibnamefont {Chen}}, \bibinfo
  {author} {\bibfnamefont {X.-L.}\ \bibnamefont {Sheng}},\ and\ \bibinfo
  {author} {\bibfnamefont {S.~A.}\ \bibnamefont {Yang}},\ }\bibfield  {title}
  {\bibinfo {title} {Switching spinless and spinful topological phases with
  projective ${PT}$ symmetry},\ }\href
  {https://doi.org/10.1103/PhysRevLett.126.196402} {\bibfield  {journal}
  {\bibinfo  {journal} {Phys. Rev. Lett.}\ }\textbf {\bibinfo {volume} {126}},\
  \bibinfo {pages} {196402} (\bibinfo {year} {2021})}\BibitemShut {NoStop}%
\bibitem [{\citenamefont {Eckardt}(2017)}]{RevModPhys.89.011004}%
  \BibitemOpen
  \bibfield  {author} {\bibinfo {author} {\bibfnamefont {A.}~\bibnamefont
  {Eckardt}},\ }\bibfield  {title} {\bibinfo {title} {Colloquium: Atomic
  quantum gases in periodically driven optical lattices},\ }\href
  {https://doi.org/10.1103/RevModPhys.89.011004} {\bibfield  {journal}
  {\bibinfo  {journal} {Rev. Mod. Phys.}\ }\textbf {\bibinfo {volume} {89}},\
  \bibinfo {pages} {011004} (\bibinfo {year} {2017})}\BibitemShut {NoStop}%
\bibitem [{\citenamefont {Bai}\ \emph {et~al.}(2021)\citenamefont {Bai},
  \citenamefont {Chen}, \citenamefont {Wu},\ and\ \citenamefont
  {An}}]{Bai_2021}%
  \BibitemOpen
  \bibfield  {author} {\bibinfo {author} {\bibfnamefont {S.-Y.}\ \bibnamefont
  {Bai}}, \bibinfo {author} {\bibfnamefont {C.}~\bibnamefont {Chen}}, \bibinfo
  {author} {\bibfnamefont {H.}~\bibnamefont {Wu}},\ and\ \bibinfo {author}
  {\bibfnamefont {J.-H.}\ \bibnamefont {An}},\ }\bibfield  {title} {\bibinfo
  {title} {Quantum control in open and periodically driven systems},\ }\href
  {https://doi.org/10.1080/23746149.2020.1870559} {\bibfield  {journal}
  {\bibinfo  {journal} {Advances in Physics: X}\ }\textbf {\bibinfo {volume}
  {6}},\ \bibinfo {pages} {1870559} (\bibinfo {year} {2021})}\BibitemShut
  {NoStop}%
\bibitem [{\citenamefont {Rudner}\ \emph {et~al.}(2013)\citenamefont {Rudner},
  \citenamefont {Lindner}, \citenamefont {Berg},\ and\ \citenamefont
  {Levin}}]{PhysRevX.3.031005}%
  \BibitemOpen
  \bibfield  {author} {\bibinfo {author} {\bibfnamefont {M.~S.}\ \bibnamefont
  {Rudner}}, \bibinfo {author} {\bibfnamefont {N.~H.}\ \bibnamefont {Lindner}},
  \bibinfo {author} {\bibfnamefont {E.}~\bibnamefont {Berg}},\ and\ \bibinfo
  {author} {\bibfnamefont {M.}~\bibnamefont {Levin}},\ }\bibfield  {title}
  {\bibinfo {title} {Anomalous edge states and the bulk-edge correspondence for
  periodically driven two-dimensional systems},\ }\href
  {https://doi.org/10.1103/PhysRevX.3.031005} {\bibfield  {journal} {\bibinfo
  {journal} {Phys. Rev. X}\ }\textbf {\bibinfo {volume} {3}},\ \bibinfo {pages}
  {031005} (\bibinfo {year} {2013})}\BibitemShut {NoStop}%
\bibitem [{\citenamefont {Slager}\ \emph {et~al.}(2024)\citenamefont {Slager},
  \citenamefont {Bouhon},\ and\ \citenamefont {Ünal}}]{floquetna}%
  \BibitemOpen
  \bibfield  {author} {\bibinfo {author} {\bibfnamefont {R.-J.}\ \bibnamefont
  {Slager}}, \bibinfo {author} {\bibfnamefont {A.}~\bibnamefont {Bouhon}},\
  and\ \bibinfo {author} {\bibfnamefont {F.~N.}\ \bibnamefont {Ünal}},\
  }\bibfield  {title} {\bibinfo {title} {Non-abelian floquet braiding and
  anomalous dirac string phase in periodically driven systems},\ }\href
  {https://doi.org/10.1038/s41467-024-45302-2} {\bibfield  {journal} {\bibinfo
  {journal} {Nature Communications}\ }\textbf {\bibinfo {volume} {15}},\
  \bibinfo {pages} {1144} (\bibinfo {year} {2024})}\BibitemShut {NoStop}%
\bibitem [{\citenamefont {McIver}\ \emph {et~al.}(2020)\citenamefont {McIver},
  \citenamefont {Schulte}, \citenamefont {Stein}, \citenamefont {Matsuyama},
  \citenamefont {Jotzu}, \citenamefont {Meier},\ and\ \citenamefont
  {Cavalleri}}]{floquetshuyun}%
  \BibitemOpen
  \bibfield  {author} {\bibinfo {author} {\bibfnamefont {J.~W.}\ \bibnamefont
  {McIver}}, \bibinfo {author} {\bibfnamefont {B.}~\bibnamefont {Schulte}},
  \bibinfo {author} {\bibfnamefont {F.-U.}\ \bibnamefont {Stein}}, \bibinfo
  {author} {\bibfnamefont {T.}~\bibnamefont {Matsuyama}}, \bibinfo {author}
  {\bibfnamefont {G.}~\bibnamefont {Jotzu}}, \bibinfo {author} {\bibfnamefont
  {G.}~\bibnamefont {Meier}},\ and\ \bibinfo {author} {\bibfnamefont
  {A.}~\bibnamefont {Cavalleri}},\ }\bibfield  {title} {\bibinfo {title}
  {Light-induced anomalous hall effect in graphene},\ }\href
  {https://doi.org/10.1038/s41567-019-0698-y} {\bibfield  {journal} {\bibinfo
  {journal} {Nature Physics}\ }\textbf {\bibinfo {volume} {16}},\ \bibinfo
  {pages} {38} (\bibinfo {year} {2020})}\BibitemShut {NoStop}%
\bibitem [{\citenamefont {Foa~Torres}\ \emph {et~al.}(2014)\citenamefont
  {Foa~Torres}, \citenamefont {Perez-Piskunow}, \citenamefont {Balseiro},\ and\
  \citenamefont {Usaj}}]{PhysRevLett.113.266801}%
  \BibitemOpen
  \bibfield  {author} {\bibinfo {author} {\bibfnamefont {L.~E.~F.}\
  \bibnamefont {Foa~Torres}}, \bibinfo {author} {\bibfnamefont {P.~M.}\
  \bibnamefont {Perez-Piskunow}}, \bibinfo {author} {\bibfnamefont {C.~A.}\
  \bibnamefont {Balseiro}},\ and\ \bibinfo {author} {\bibfnamefont
  {G.}~\bibnamefont {Usaj}},\ }\bibfield  {title} {\bibinfo {title}
  {Multiterminal conductance of a floquet topological insulator},\ }\href
  {https://doi.org/10.1103/PhysRevLett.113.266801} {\bibfield  {journal}
  {\bibinfo  {journal} {Phys. Rev. Lett.}\ }\textbf {\bibinfo {volume} {113}},\
  \bibinfo {pages} {266801} (\bibinfo {year} {2014})}\BibitemShut {NoStop}%
\bibitem [{\citenamefont {Wu}\ and\ \citenamefont
  {An}(2023)}]{PhysRevB.107.235132}%
  \BibitemOpen
  \bibfield  {author} {\bibinfo {author} {\bibfnamefont {H.}~\bibnamefont
  {Wu}}\ and\ \bibinfo {author} {\bibfnamefont {J.-H.}\ \bibnamefont {An}},\
  }\bibfield  {title} {\bibinfo {title} {Hybrid-order topological odd-parity
  superconductors via floquet engineering},\ }\href
  {https://doi.org/10.1103/PhysRevB.107.235132} {\bibfield  {journal} {\bibinfo
   {journal} {Phys. Rev. B}\ }\textbf {\bibinfo {volume} {107}},\ \bibinfo
  {pages} {235132} (\bibinfo {year} {2023})}\BibitemShut {NoStop}%
\bibitem [{\citenamefont {Chen}\ \emph {et~al.}(2015)\citenamefont {Chen},
  \citenamefont {An}, \citenamefont {Luo}, \citenamefont {Sun},\ and\
  \citenamefont {Oh}}]{PhysRevA.91.052122}%
  \BibitemOpen
  \bibfield  {author} {\bibinfo {author} {\bibfnamefont {C.}~\bibnamefont
  {Chen}}, \bibinfo {author} {\bibfnamefont {J.-H.}\ \bibnamefont {An}},
  \bibinfo {author} {\bibfnamefont {H.-G.}\ \bibnamefont {Luo}}, \bibinfo
  {author} {\bibfnamefont {C.~P.}\ \bibnamefont {Sun}},\ and\ \bibinfo {author}
  {\bibfnamefont {C.~H.}\ \bibnamefont {Oh}},\ }\bibfield  {title} {\bibinfo
  {title} {Floquet control of quantum dissipation in spin chains},\ }\href
  {https://doi.org/10.1103/PhysRevA.91.052122} {\bibfield  {journal} {\bibinfo
  {journal} {Phys. Rev. A}\ }\textbf {\bibinfo {volume} {91}},\ \bibinfo
  {pages} {052122} (\bibinfo {year} {2015})}\BibitemShut {NoStop}%
\bibitem [{\citenamefont {Kundu}\ \emph {et~al.}(2014)\citenamefont {Kundu},
  \citenamefont {Fertig},\ and\ \citenamefont
  {Seradjeh}}]{PhysRevLett.113.236803}%
  \BibitemOpen
  \bibfield  {author} {\bibinfo {author} {\bibfnamefont {A.}~\bibnamefont
  {Kundu}}, \bibinfo {author} {\bibfnamefont {H.~A.}\ \bibnamefont {Fertig}},\
  and\ \bibinfo {author} {\bibfnamefont {B.}~\bibnamefont {Seradjeh}},\
  }\bibfield  {title} {\bibinfo {title} {Effective theory of {F}loquet
  topological transitions},\ }\href
  {https://doi.org/10.1103/PhysRevLett.113.236803} {\bibfield  {journal}
  {\bibinfo  {journal} {Phys. Rev. Lett.}\ }\textbf {\bibinfo {volume} {113}},\
  \bibinfo {pages} {236803} (\bibinfo {year} {2014})}\BibitemShut {NoStop}%
\bibitem [{\citenamefont {Rodriguez-Vega}\ \emph {et~al.}(2019)\citenamefont
  {Rodriguez-Vega}, \citenamefont {Kumar},\ and\ \citenamefont
  {Seradjeh}}]{PhysRevB.100.085138}%
  \BibitemOpen
  \bibfield  {author} {\bibinfo {author} {\bibfnamefont {M.}~\bibnamefont
  {Rodriguez-Vega}}, \bibinfo {author} {\bibfnamefont {A.}~\bibnamefont
  {Kumar}},\ and\ \bibinfo {author} {\bibfnamefont {B.}~\bibnamefont
  {Seradjeh}},\ }\bibfield  {title} {\bibinfo {title} {Higher-order {F}loquet
  topological phases with corner and bulk bound states},\ }\href
  {https://doi.org/10.1103/PhysRevB.100.085138} {\bibfield  {journal} {\bibinfo
   {journal} {Phys. Rev. B}\ }\textbf {\bibinfo {volume} {100}},\ \bibinfo
  {pages} {085138} (\bibinfo {year} {2019})}\BibitemShut {NoStop}%
\bibitem [{\citenamefont {Wu}\ and\ \citenamefont
  {An}(2020)}]{PhysRevB.102.041119}%
  \BibitemOpen
  \bibfield  {author} {\bibinfo {author} {\bibfnamefont {H.}~\bibnamefont
  {Wu}}\ and\ \bibinfo {author} {\bibfnamefont {J.-H.}\ \bibnamefont {An}},\
  }\bibfield  {title} {\bibinfo {title} {Floquet topological phases of
  non-{H}ermitian systems},\ }\href
  {https://doi.org/10.1103/PhysRevB.102.041119} {\bibfield  {journal} {\bibinfo
   {journal} {Phys. Rev. B}\ }\textbf {\bibinfo {volume} {102}},\ \bibinfo
  {pages} {041119(R)} (\bibinfo {year} {2020})}\BibitemShut {NoStop}%
\bibitem [{\citenamefont {Rodriguez-Vega}\ \emph {et~al.}(2020)\citenamefont
  {Rodriguez-Vega}, \citenamefont {Vogl},\ and\ \citenamefont
  {Fiete}}]{PhysRevResearch.2.033494}%
  \BibitemOpen
  \bibfield  {author} {\bibinfo {author} {\bibfnamefont {M.}~\bibnamefont
  {Rodriguez-Vega}}, \bibinfo {author} {\bibfnamefont {M.}~\bibnamefont
  {Vogl}},\ and\ \bibinfo {author} {\bibfnamefont {G.~A.}\ \bibnamefont
  {Fiete}},\ }\bibfield  {title} {\bibinfo {title} {Floquet engineering of
  twisted double bilayer graphene},\ }\href
  {https://doi.org/10.1103/PhysRevResearch.2.033494} {\bibfield  {journal}
  {\bibinfo  {journal} {Phys. Rev. Res.}\ }\textbf {\bibinfo {volume} {2}},\
  \bibinfo {pages} {033494} (\bibinfo {year} {2020})}\BibitemShut {NoStop}%
\bibitem [{\citenamefont {Yao}\ \emph {et~al.}(2017)\citenamefont {Yao},
  \citenamefont {Yan},\ and\ \citenamefont {Wang}}]{PhysRevB.96.195303}%
  \BibitemOpen
  \bibfield  {author} {\bibinfo {author} {\bibfnamefont {S.}~\bibnamefont
  {Yao}}, \bibinfo {author} {\bibfnamefont {Z.}~\bibnamefont {Yan}},\ and\
  \bibinfo {author} {\bibfnamefont {Z.}~\bibnamefont {Wang}},\ }\bibfield
  {title} {\bibinfo {title} {Topological invariants of {F}loquet systems:
  General formulation, special properties, and {F}loquet topological defects},\
  }\href {https://doi.org/10.1103/PhysRevB.96.195303} {\bibfield  {journal}
  {\bibinfo  {journal} {Phys. Rev. B}\ }\textbf {\bibinfo {volume} {96}},\
  \bibinfo {pages} {195303} (\bibinfo {year} {2017})}\BibitemShut {NoStop}%
\bibitem [{\citenamefont {Wang}\ \emph {et~al.}(2021)\citenamefont {Wang},
  \citenamefont {Wu},\ and\ \citenamefont {An}}]{PhysRevB.104.205117}%
  \BibitemOpen
  \bibfield  {author} {\bibinfo {author} {\bibfnamefont {B.-Q.}\ \bibnamefont
  {Wang}}, \bibinfo {author} {\bibfnamefont {H.}~\bibnamefont {Wu}},\ and\
  \bibinfo {author} {\bibfnamefont {J.-H.}\ \bibnamefont {An}},\ }\bibfield
  {title} {\bibinfo {title} {Engineering exotic second-order topological
  semimetals by periodic driving},\ }\href
  {https://doi.org/10.1103/PhysRevB.104.205117} {\bibfield  {journal} {\bibinfo
   {journal} {Phys. Rev. B}\ }\textbf {\bibinfo {volume} {104}},\ \bibinfo
  {pages} {205117} (\bibinfo {year} {2021})}\BibitemShut {NoStop}%
\bibitem [{\citenamefont {Zhang}\ and\ \citenamefont
  {Yang}(2020)}]{zhang2020theoryanomalousfloquethigherorder}%
  \BibitemOpen
  \bibfield  {author} {\bibinfo {author} {\bibfnamefont {R.-X.}\ \bibnamefont
  {Zhang}}\ and\ \bibinfo {author} {\bibfnamefont {Z.-C.}\ \bibnamefont
  {Yang}},\ }\href {https://arxiv.org/abs/2010.07945} {\bibinfo {title} {Theory
  of anomalous floquet higher-order topology: Classification, characterization,
  and bulk-boundary correspondence}} (\bibinfo {year} {2020}),\ \Eprint
  {https://arxiv.org/abs/2010.07945} {arXiv:2010.07945 [cond-mat.mes-{H}all]}
  \BibitemShut {NoStop}%
\bibitem [{\citenamefont {Ezawa}(2020)}]{PhysRevB.102.121405}%
  \BibitemOpen
  \bibfield  {author} {\bibinfo {author} {\bibfnamefont {M.}~\bibnamefont
  {Ezawa}},\ }\bibfield  {title} {\bibinfo {title} {Edge-corner correspondence:
  Boundary-obstructed topological phases with chiral symmetry},\ }\href
  {https://doi.org/10.1103/PhysRevB.102.121405} {\bibfield  {journal} {\bibinfo
   {journal} {Phys. Rev. B}\ }\textbf {\bibinfo {volume} {102}},\ \bibinfo
  {pages} {121405} (\bibinfo {year} {2020})}\BibitemShut {NoStop}%
\bibitem [{\citenamefont {Jackiw}\ and\ \citenamefont
  {Rebbi}(1976)}]{PhysRevD.13.3398}%
  \BibitemOpen
  \bibfield  {author} {\bibinfo {author} {\bibfnamefont {R.}~\bibnamefont
  {Jackiw}}\ and\ \bibinfo {author} {\bibfnamefont {C.}~\bibnamefont {Rebbi}},\
  }\bibfield  {title} {\bibinfo {title} {Solitons with fermion number
  \textonehalf{}},\ }\href {https://doi.org/10.1103/PhysRevD.13.3398}
  {\bibfield  {journal} {\bibinfo  {journal} {Phys. Rev. D}\ }\textbf {\bibinfo
  {volume} {13}},\ \bibinfo {pages} {3398} (\bibinfo {year}
  {1976})}\BibitemShut {NoStop}%
\bibitem [{\citenamefont {Ma}\ \emph {et~al.}(2024)\citenamefont {Ma},
  \citenamefont {Pu}, \citenamefont {Ye}, \citenamefont {Lu}, \citenamefont
  {Huang}, \citenamefont {Ke}, \citenamefont {He}, \citenamefont {Deng},\ and\
  \citenamefont {Liu}}]{PhysRevLett.132.066601}%
  \BibitemOpen
  \bibfield  {author} {\bibinfo {author} {\bibfnamefont {Q.}~\bibnamefont
  {Ma}}, \bibinfo {author} {\bibfnamefont {Z.}~\bibnamefont {Pu}}, \bibinfo
  {author} {\bibfnamefont {L.}~\bibnamefont {Ye}}, \bibinfo {author}
  {\bibfnamefont {J.}~\bibnamefont {Lu}}, \bibinfo {author} {\bibfnamefont
  {X.}~\bibnamefont {Huang}}, \bibinfo {author} {\bibfnamefont
  {M.}~\bibnamefont {Ke}}, \bibinfo {author} {\bibfnamefont {H.}~\bibnamefont
  {He}}, \bibinfo {author} {\bibfnamefont {W.}~\bibnamefont {Deng}},\ and\
  \bibinfo {author} {\bibfnamefont {Z.}~\bibnamefont {Liu}},\ }\bibfield
  {title} {\bibinfo {title} {Observation of higher-order nodal-line semimetal
  in phononic crystals},\ }\href
  {https://doi.org/10.1103/PhysRevLett.132.066601} {\bibfield  {journal}
  {\bibinfo  {journal} {Phys. Rev. Lett.}\ }\textbf {\bibinfo {volume} {132}},\
  \bibinfo {pages} {066601} (\bibinfo {year} {2024})}\BibitemShut {NoStop}%
\bibitem [{\citenamefont {Xiang}\ \emph {et~al.}(2024)\citenamefont {Xiang},
  \citenamefont {Peng}, \citenamefont {Gao}, \citenamefont {Wu}, \citenamefont
  {Wu}, \citenamefont {Chen}, \citenamefont {Ni},\ and\ \citenamefont
  {Zhu}}]{PhysRevLett.132.197202}%
  \BibitemOpen
  \bibfield  {author} {\bibinfo {author} {\bibfnamefont {X.}~\bibnamefont
  {Xiang}}, \bibinfo {author} {\bibfnamefont {Y.-G.}\ \bibnamefont {Peng}},
  \bibinfo {author} {\bibfnamefont {F.}~\bibnamefont {Gao}}, \bibinfo {author}
  {\bibfnamefont {X.}~\bibnamefont {Wu}}, \bibinfo {author} {\bibfnamefont
  {P.}~\bibnamefont {Wu}}, \bibinfo {author} {\bibfnamefont {Z.}~\bibnamefont
  {Chen}}, \bibinfo {author} {\bibfnamefont {X.}~\bibnamefont {Ni}},\ and\
  \bibinfo {author} {\bibfnamefont {X.-F.}\ \bibnamefont {Zhu}},\ }\bibfield
  {title} {\bibinfo {title} {Demonstration of acoustic higher-order topological
  {S}tiefel-{W}hitney semimetal},\ }\href
  {https://doi.org/10.1103/PhysRevLett.132.197202} {\bibfield  {journal}
  {\bibinfo  {journal} {Phys. Rev. Lett.}\ }\textbf {\bibinfo {volume} {132}},\
  \bibinfo {pages} {197202} (\bibinfo {year} {2024})}\BibitemShut {NoStop}%
\bibitem [{\citenamefont {Luo}\ \emph {et~al.}(2021)\citenamefont {Luo},
  \citenamefont {Wang}, \citenamefont {Lin}, \citenamefont {Jiang},
  \citenamefont {Wu}, \citenamefont {Li},\ and\ \citenamefont
  {Jiang}}]{Luo_2021}%
  \BibitemOpen
  \bibfield  {author} {\bibinfo {author} {\bibfnamefont {L.}~\bibnamefont
  {Luo}}, \bibinfo {author} {\bibfnamefont {H.-X.}\ \bibnamefont {Wang}},
  \bibinfo {author} {\bibfnamefont {Z.-K.}\ \bibnamefont {Lin}}, \bibinfo
  {author} {\bibfnamefont {B.}~\bibnamefont {Jiang}}, \bibinfo {author}
  {\bibfnamefont {Y.}~\bibnamefont {Wu}}, \bibinfo {author} {\bibfnamefont
  {F.}~\bibnamefont {Li}},\ and\ \bibinfo {author} {\bibfnamefont {J.-H.}\
  \bibnamefont {Jiang}},\ }\bibfield  {title} {\bibinfo {title} {Observation of
  a phononic higher-order {W}eyl semimetal},\ }\href
  {https://doi.org/10.1038/s41563-021-00985-6} {\bibfield  {journal} {\bibinfo
  {journal} {Nature Materials}\ }\textbf {\bibinfo {volume} {20}},\ \bibinfo
  {pages} {794–799} (\bibinfo {year} {2021})}\BibitemShut {NoStop}%
\bibitem [{\citenamefont {Xue}\ \emph {et~al.}(2023)\citenamefont {Xue},
  \citenamefont {Chen}, \citenamefont {Cheng}, \citenamefont {Dai},
  \citenamefont {Long}, \citenamefont {Zhao},\ and\ \citenamefont
  {Zhang}}]{Xue_2023}%
  \BibitemOpen
  \bibfield  {author} {\bibinfo {author} {\bibfnamefont {H.}~\bibnamefont
  {Xue}}, \bibinfo {author} {\bibfnamefont {Z.~Y.}\ \bibnamefont {Chen}},
  \bibinfo {author} {\bibfnamefont {Z.}~\bibnamefont {Cheng}}, \bibinfo
  {author} {\bibfnamefont {J.~X.}\ \bibnamefont {Dai}}, \bibinfo {author}
  {\bibfnamefont {Y.}~\bibnamefont {Long}}, \bibinfo {author} {\bibfnamefont
  {Y.~X.}\ \bibnamefont {Zhao}},\ and\ \bibinfo {author} {\bibfnamefont
  {B.}~\bibnamefont {Zhang}},\ }\bibfield  {title} {\bibinfo {title}
  {Stiefel-{W}hitney topological charges in a three-dimensional acoustic
  nodal-line crystal},\ }\href {https://doi.org/10.1038/s41467-023-40252-7}
  {\bibfield  {journal} {\bibinfo  {journal} {Nature Communications}\ }\textbf
  {\bibinfo {volume} {14}},\ \bibinfo {pages} {4563} (\bibinfo {year}
  {2023})}\BibitemShut {NoStop}%
\bibitem [{\citenamefont {Pan}\ \emph {et~al.}(2023)\citenamefont {Pan},
  \citenamefont {Cui}, \citenamefont {Chen}, \citenamefont {Chen},
  \citenamefont {Zhang}, \citenamefont {Ren}, \citenamefont {Han},
  \citenamefont {Li}, \citenamefont {Li}, \citenamefont {Yu}, \citenamefont
  {Chen},\ and\ \citenamefont {Yang}}]{Pan2023}%
  \BibitemOpen
  \bibfield  {author} {\bibinfo {author} {\bibfnamefont {Y.}~\bibnamefont
  {Pan}}, \bibinfo {author} {\bibfnamefont {C.}~\bibnamefont {Cui}}, \bibinfo
  {author} {\bibfnamefont {Q.}~\bibnamefont {Chen}}, \bibinfo {author}
  {\bibfnamefont {F.}~\bibnamefont {Chen}}, \bibinfo {author} {\bibfnamefont
  {L.}~\bibnamefont {Zhang}}, \bibinfo {author} {\bibfnamefont
  {Y.}~\bibnamefont {Ren}}, \bibinfo {author} {\bibfnamefont {N.}~\bibnamefont
  {Han}}, \bibinfo {author} {\bibfnamefont {W.}~\bibnamefont {Li}}, \bibinfo
  {author} {\bibfnamefont {X.}~\bibnamefont {Li}}, \bibinfo {author}
  {\bibfnamefont {Z.-M.}\ \bibnamefont {Yu}}, \bibinfo {author} {\bibfnamefont
  {H.}~\bibnamefont {Chen}},\ and\ \bibinfo {author} {\bibfnamefont
  {Y.}~\bibnamefont {Yang}},\ }\bibfield  {title} {\bibinfo {title} {Real
  higher-order weyl photonic crystal},\ }\href
  {https://doi.org/10.1038/s41467-023-42457-2} {\bibfield  {journal} {\bibinfo
  {journal} {Nature Communications}\ }\textbf {\bibinfo {volume} {14}},\
  \bibinfo {pages} {6636} (\bibinfo {year} {2023})}\BibitemShut {NoStop}%
\bibitem [{\citenamefont {Meinert}\ \emph {et~al.}(2016)\citenamefont
  {Meinert}, \citenamefont {Mark}, \citenamefont {Lauber}, \citenamefont
  {Daley},\ and\ \citenamefont {N\"agerl}}]{PhysRevLett.116.205301}%
  \BibitemOpen
  \bibfield  {author} {\bibinfo {author} {\bibfnamefont {F.}~\bibnamefont
  {Meinert}}, \bibinfo {author} {\bibfnamefont {M.~J.}\ \bibnamefont {Mark}},
  \bibinfo {author} {\bibfnamefont {K.}~\bibnamefont {Lauber}}, \bibinfo
  {author} {\bibfnamefont {A.~J.}\ \bibnamefont {Daley}},\ and\ \bibinfo
  {author} {\bibfnamefont {H.-C.}\ \bibnamefont {N\"agerl}},\ }\bibfield
  {title} {\bibinfo {title} {Floquet engineering of correlated tunneling in the
  {B}ose-{H}ubbard model with ultracold atoms},\ }\href
  {https://doi.org/10.1103/PhysRevLett.116.205301} {\bibfield  {journal}
  {\bibinfo  {journal} {Phys. Rev. Lett.}\ }\textbf {\bibinfo {volume} {116}},\
  \bibinfo {pages} {205301} (\bibinfo {year} {2016})}\BibitemShut {NoStop}%
\bibitem [{\citenamefont {Zhang}\ \emph {et~al.}(2023)\citenamefont {Zhang},
  \citenamefont {Yi}, \citenamefont {Zhang}, \citenamefont {Jiao},
  \citenamefont {Shi}, \citenamefont {Yuan}, \citenamefont {Zhang},
  \citenamefont {Liu}, \citenamefont {Chen},\ and\ \citenamefont
  {Pan}}]{PhysRevLett.130.043201}%
  \BibitemOpen
  \bibfield  {author} {\bibinfo {author} {\bibfnamefont {J.-Y.}\ \bibnamefont
  {Zhang}}, \bibinfo {author} {\bibfnamefont {C.-R.}\ \bibnamefont {Yi}},
  \bibinfo {author} {\bibfnamefont {L.}~\bibnamefont {Zhang}}, \bibinfo
  {author} {\bibfnamefont {R.-H.}\ \bibnamefont {Jiao}}, \bibinfo {author}
  {\bibfnamefont {K.-Y.}\ \bibnamefont {Shi}}, \bibinfo {author} {\bibfnamefont
  {H.}~\bibnamefont {Yuan}}, \bibinfo {author} {\bibfnamefont {W.}~\bibnamefont
  {Zhang}}, \bibinfo {author} {\bibfnamefont {X.-J.}\ \bibnamefont {Liu}},
  \bibinfo {author} {\bibfnamefont {S.}~\bibnamefont {Chen}},\ and\ \bibinfo
  {author} {\bibfnamefont {J.-W.}\ \bibnamefont {Pan}},\ }\bibfield  {title}
  {\bibinfo {title} {Tuning anomalous {F}loquet topological bands with
  ultracold atoms},\ }\href {https://doi.org/10.1103/PhysRevLett.130.043201}
  {\bibfield  {journal} {\bibinfo  {journal} {Phys. Rev. Lett.}\ }\textbf
  {\bibinfo {volume} {130}},\ \bibinfo {pages} {043201} (\bibinfo {year}
  {2023})}\BibitemShut {NoStop}%
\bibitem [{\citenamefont {Roushan}\ \emph {et~al.}(2017)\citenamefont
  {Roushan}, \citenamefont {Neill}, \citenamefont {Megrant}, \citenamefont
  {Chen}, \citenamefont {Babbush}, \citenamefont {Barends}, \citenamefont
  {Campbell}, \citenamefont {Chen}, \citenamefont {Chiaro}, \citenamefont
  {Dunsworth}, \citenamefont {Fowler}, \citenamefont {Jeffrey}, \citenamefont
  {Kelly}, \citenamefont {Lucero}, \citenamefont {Mutus}, \citenamefont {OHuo
  Heng~alley}, \citenamefont {Neeley}, \citenamefont {Quintana}, \citenamefont
  {Sank}, \citenamefont {Vainsencher}, \citenamefont {Wenner}, \citenamefont
  {White}, \citenamefont {Kapit}, \citenamefont {Neven},\ and\ \citenamefont
  {Martinis}}]{Roushan2017}%
  \BibitemOpen
  \bibfield  {author} {\bibinfo {author} {\bibfnamefont {P.}~\bibnamefont
  {Roushan}}, \bibinfo {author} {\bibfnamefont {C.}~\bibnamefont {Neill}},
  \bibinfo {author} {\bibfnamefont {A.}~\bibnamefont {Megrant}}, \bibinfo
  {author} {\bibfnamefont {Y.}~\bibnamefont {Chen}}, \bibinfo {author}
  {\bibfnamefont {R.}~\bibnamefont {Babbush}}, \bibinfo {author} {\bibfnamefont
  {R.}~\bibnamefont {Barends}}, \bibinfo {author} {\bibfnamefont
  {B.}~\bibnamefont {Campbell}}, \bibinfo {author} {\bibfnamefont
  {Z.}~\bibnamefont {Chen}}, \bibinfo {author} {\bibfnamefont {B.}~\bibnamefont
  {Chiaro}}, \bibinfo {author} {\bibfnamefont {A.}~\bibnamefont {Dunsworth}},
  \bibinfo {author} {\bibfnamefont {A.}~\bibnamefont {Fowler}}, \bibinfo
  {author} {\bibfnamefont {E.}~\bibnamefont {Jeffrey}}, \bibinfo {author}
  {\bibfnamefont {J.}~\bibnamefont {Kelly}}, \bibinfo {author} {\bibfnamefont
  {E.}~\bibnamefont {Lucero}}, \bibinfo {author} {\bibfnamefont
  {J.}~\bibnamefont {Mutus}}, \bibinfo {author} {\bibfnamefont {P.~J.~J.}\
  \bibnamefont {OHuo Heng~alley}}, \bibinfo {author} {\bibfnamefont
  {M.}~\bibnamefont {Neeley}}, \bibinfo {author} {\bibfnamefont
  {C.}~\bibnamefont {Quintana}}, \bibinfo {author} {\bibfnamefont
  {D.}~\bibnamefont {Sank}}, \bibinfo {author} {\bibfnamefont {A.}~\bibnamefont
  {Vainsencher}}, \bibinfo {author} {\bibfnamefont {J.}~\bibnamefont {Wenner}},
  \bibinfo {author} {\bibfnamefont {T.}~\bibnamefont {White}}, \bibinfo
  {author} {\bibfnamefont {E.}~\bibnamefont {Kapit}}, \bibinfo {author}
  {\bibfnamefont {H.}~\bibnamefont {Neven}},\ and\ \bibinfo {author}
  {\bibfnamefont {J.}~\bibnamefont {Martinis}},\ }\bibfield  {title} {\bibinfo
  {title} {Chiral ground-state currents of interacting photons in a synthetic
  magnetic field},\ }\href {https://doi.org/10.1038/nphys3930} {\bibfield
  {journal} {\bibinfo  {journal} {Nature Physics}\ }\textbf {\bibinfo {volume}
  {13}},\ \bibinfo {pages} {146} (\bibinfo {year} {2017})}\BibitemShut
  {NoStop}%
\bibitem [{\citenamefont {Rechtsman}\ \emph {et~al.}(2013)\citenamefont
  {Rechtsman}, \citenamefont {Zeuner}, \citenamefont {Plotnik}, \citenamefont
  {Lumer}, \citenamefont {Podolsky}, \citenamefont {Dreisow}, \citenamefont
  {Nolte}, \citenamefont {Segev},\ and\ \citenamefont
  {Szameit}}]{Rechtsman2013}%
  \BibitemOpen
  \bibfield  {author} {\bibinfo {author} {\bibfnamefont {M.~C.}\ \bibnamefont
  {Rechtsman}}, \bibinfo {author} {\bibfnamefont {J.~M.}\ \bibnamefont
  {Zeuner}}, \bibinfo {author} {\bibfnamefont {Y.}~\bibnamefont {Plotnik}},
  \bibinfo {author} {\bibfnamefont {Y.}~\bibnamefont {Lumer}}, \bibinfo
  {author} {\bibfnamefont {D.}~\bibnamefont {Podolsky}}, \bibinfo {author}
  {\bibfnamefont {F.}~\bibnamefont {Dreisow}}, \bibinfo {author} {\bibfnamefont
  {S.}~\bibnamefont {Nolte}}, \bibinfo {author} {\bibfnamefont
  {M.}~\bibnamefont {Segev}},\ and\ \bibinfo {author} {\bibfnamefont
  {A.}~\bibnamefont {Szameit}},\ }\bibfield  {title} {\bibinfo {title}
  {Photonic {F}loquet topological insulators},\ }\href
  {https://doi.org/10.1038/nature12066} {\bibfield  {journal} {\bibinfo
  {journal} {Nature}\ }\textbf {\bibinfo {volume} {496}},\ \bibinfo {pages}
  {196} (\bibinfo {year} {2013})}\BibitemShut {NoStop}%
\bibitem [{\citenamefont {Cheng}\ \emph {et~al.}(2019)\citenamefont {Cheng},
  \citenamefont {Pan}, \citenamefont {Wang}, \citenamefont {Zhang},
  \citenamefont {Yu}, \citenamefont {Gover}, \citenamefont {Zhang},
  \citenamefont {Li}, \citenamefont {Zhou},\ and\ \citenamefont
  {Zhu}}]{PhysRevLett.122.173901}%
  \BibitemOpen
  \bibfield  {author} {\bibinfo {author} {\bibfnamefont {Q.}~\bibnamefont
  {Cheng}}, \bibinfo {author} {\bibfnamefont {Y.}~\bibnamefont {Pan}}, \bibinfo
  {author} {\bibfnamefont {H.}~\bibnamefont {Wang}}, \bibinfo {author}
  {\bibfnamefont {C.}~\bibnamefont {Zhang}}, \bibinfo {author} {\bibfnamefont
  {D.}~\bibnamefont {Yu}}, \bibinfo {author} {\bibfnamefont {A.}~\bibnamefont
  {Gover}}, \bibinfo {author} {\bibfnamefont {H.}~\bibnamefont {Zhang}},
  \bibinfo {author} {\bibfnamefont {T.}~\bibnamefont {Li}}, \bibinfo {author}
  {\bibfnamefont {L.}~\bibnamefont {Zhou}},\ and\ \bibinfo {author}
  {\bibfnamefont {S.}~\bibnamefont {Zhu}},\ }\bibfield  {title} {\bibinfo
  {title} {Observation of anomalous $\ensuremath{\pi}$ modes in photonic
  {F}loquet engineering},\ }\href
  {https://doi.org/10.1103/PhysRevLett.122.173901} {\bibfield  {journal}
  {\bibinfo  {journal} {Phys. Rev. Lett.}\ }\textbf {\bibinfo {volume} {122}},\
  \bibinfo {pages} {173901} (\bibinfo {year} {2019})}\BibitemShut {NoStop}%
\bibitem [{\citenamefont {Lin}\ \emph {et~al.}(2024)\citenamefont {Lin},
  \citenamefont {Song}, \citenamefont {Wang}, \citenamefont {Xin},
  \citenamefont {Sun}, \citenamefont {Wu}, \citenamefont {Huang}, \citenamefont
  {Zhu}, \citenamefont {Jiang},\ and\ \citenamefont
  {Li}}]{PhysRevLett.133.073803}%
  \BibitemOpen
  \bibfield  {author} {\bibinfo {author} {\bibfnamefont {Z.}~\bibnamefont
  {Lin}}, \bibinfo {author} {\bibfnamefont {W.}~\bibnamefont {Song}}, \bibinfo
  {author} {\bibfnamefont {L.-W.}\ \bibnamefont {Wang}}, \bibinfo {author}
  {\bibfnamefont {H.}~\bibnamefont {Xin}}, \bibinfo {author} {\bibfnamefont
  {J.}~\bibnamefont {Sun}}, \bibinfo {author} {\bibfnamefont {S.}~\bibnamefont
  {Wu}}, \bibinfo {author} {\bibfnamefont {C.}~\bibnamefont {Huang}}, \bibinfo
  {author} {\bibfnamefont {S.}~\bibnamefont {Zhu}}, \bibinfo {author}
  {\bibfnamefont {J.-H.}\ \bibnamefont {Jiang}},\ and\ \bibinfo {author}
  {\bibfnamefont {T.}~\bibnamefont {Li}},\ }\bibfield  {title} {\bibinfo
  {title} {Observation of topological transition in {F}loquet non-{H}ermitian
  skin effects in silicon photonics},\ }\href
  {https://doi.org/10.1103/PhysRevLett.133.073803} {\bibfield  {journal}
  {\bibinfo  {journal} {Phys. Rev. Lett.}\ }\textbf {\bibinfo {volume} {133}},\
  \bibinfo {pages} {073803} (\bibinfo {year} {2024})}\BibitemShut {NoStop}%
\bibitem [{\citenamefont {Cheng}\ \emph {et~al.}(2022)\citenamefont {Cheng},
  \citenamefont {Bomantara}, \citenamefont {Xue}, \citenamefont {Zhu},
  \citenamefont {Gong},\ and\ \citenamefont {Zhang}}]{PhysRevLett.129.254301}%
  \BibitemOpen
  \bibfield  {author} {\bibinfo {author} {\bibfnamefont {Z.}~\bibnamefont
  {Cheng}}, \bibinfo {author} {\bibfnamefont {R.~W.}\ \bibnamefont
  {Bomantara}}, \bibinfo {author} {\bibfnamefont {H.}~\bibnamefont {Xue}},
  \bibinfo {author} {\bibfnamefont {W.}~\bibnamefont {Zhu}}, \bibinfo {author}
  {\bibfnamefont {J.}~\bibnamefont {Gong}},\ and\ \bibinfo {author}
  {\bibfnamefont {B.}~\bibnamefont {Zhang}},\ }\bibfield  {title} {\bibinfo
  {title} {Observation of $\ensuremath{\pi}/2$ modes in an acoustic {F}loquet
  system},\ }\href {https://doi.org/10.1103/PhysRevLett.129.254301} {\bibfield
  {journal} {\bibinfo  {journal} {Phys. Rev. Lett.}\ }\textbf {\bibinfo
  {volume} {129}},\ \bibinfo {pages} {254301} (\bibinfo {year}
  {2022})}\BibitemShut {NoStop}%
\end{thebibliography}%
\end{document}